\documentclass[twocolumn,notitlepage,prl,superscriptaddress,longbibliography]{revtex4-2}

\usepackage{graphicx}
\usepackage{epstopdf}
\usepackage{color}
\usepackage{xcolor}
\usepackage{physics}
\usepackage{amsthm,amsmath,amssymb}
\usepackage[colorlinks=true, urlcolor=blue, linkcolor=blue, citecolor=blue, pdftex]{hyperref}
\usepackage{array}
\usepackage{multirow}
\usepackage{babel}
\usepackage{textcomp}
\usepackage{float}
\newcommand{\diag}[2]{\vcenter{\hbox{\includegraphics[height=#2]{#1}}}}

\begin{document}
\title{Topological Chiral Superconductivity in the Triangular-Lattice Hofstadter-Hubbard Model}

\author{Feng Chen}
\thanks{These authors contributed equally to this work.}
\affiliation{The Hamburg Centre for Ultrafast Imaging and Institute for Quantum Physics, University of Hamburg, Luruper Chaussee 149,
Hamburg 22761, Germany}
\affiliation{Department of Physics and Astronomy, California State University Northridge, Northridge, California 91330, USA}

\author{Wen O. Wang}
\thanks{These authors contributed equally to this work.}
\affiliation{Kavli Institute for Theoretical Physics, University of California, Santa Barbara, California 93106-4030, USA}

\author{Jia-Xin Zhang}
\email{zhangjx.phy@gmail.com}
\affiliation{French American Center for Theoretical Science, CNRS, KITP, Santa Barbara, California 93106-4030, USA}
\affiliation{Kavli Institute for Theoretical Physics, University of California, Santa Barbara, California 93106-4030, USA}

\author{Leon Balents}
\email{balents@spinsandelectrons.com}
\affiliation{Kavli Institute for Theoretical Physics, University of California, Santa Barbara, California 93106-4030, USA}
\affiliation{French American Center for Theoretical Science, CNRS, KITP, Santa Barbara, California 93106-4030, USA}
\affiliation{Canadian Institute for Advanced Research, Toronto, Ontario, M5G 1M1 Canada}

\author{D. N. Sheng}  
\email{donna.sheng1@csun.edu}
\affiliation{Department of Physics and Astronomy, California State University Northridge, Northridge, California 91330, USA}

\begin{abstract}
    Moir\'e materials provide exciting platforms for studying the interplay of strong electronic correlation and large magnetic flux effects. We study the lightly doped Hofstadter-Hubbard model 
    on a triangular lattice through large-scale density matrix renormalization group and 
    determinantal quantum Monte Carlo   simulations.  We find strong evidence for a robust chiral superconducting (SC) phase with  dominant power-law  pairing  correlations 
    and a quantized spin Chern number.  The SC phase
    emerges at very weak interaction and grows stronger at intermediate interaction strengths ($U$) for a wide range of hole doping. We also discuss the possible distinct nature of the normal state in different $U$ regimes. Our work provides theoretical insights into  the emergence of topological superconductivity from doping topological Chern bands or magnetic flux induced chiral spin liquid states of Moir\'e materials. 
\end{abstract}

\maketitle

{\it Introduction.---}
Understanding the mechanism of unconventional superconductivity (SC) in doped Mott insulators is a central topic in condensed matter physics~\cite{keimer2015quantum,lee2006doping}. Ever since Anderson's proposal of the resonating valence bond state as the undoped parent state of high-temperature SC in cuprates~\cite{anderson1987resonating}, quantum spin liquid and high-$T_c$ SC have been naturally connected~\cite{fabrizioSuperconductivityDopingSpinliquid1996,konikDopedSpinLiquid2006,rokhsarPairingDopedSpin1993,kellyElectronDopingKagome2016,senthilCupratesDoped$U1$2005,sigristSuperconductivityQuasionedimensionalSpin1994}. 
A particular class of quantum spin liquid that breaks time-reversal symmetry (TRS) and exhibits chiral edge modes is the chiral spin liquid (CSL)~\cite{kalmeyer1987equivalence}, which was predicted to give rise to $d+id$-wave chiral SC through the condensation of fractional quasiparticles, known as anyon superconductivity~\cite{laughlin1988,wen1989chiral,lee1989}.

Two dimensional spin systems on geometric frustrated lattices are promising hosts for the CSL~\cite{balents2010spin}. Numerical studies have established its existence in both the spin-1/2 kagome antiferromagnet~\cite{he2014,gong2014,bauer2014,gong2015}, and the half-filled triangular-lattice Hubbard model near the metal-insulator transition\cite{szasz2020chiral,chen2021quantum,wietek2021,zhou2022}. However,   non-SC charge density wave (CDW)~\cite{peng2021doping} or chiral metal phases~\cite{zhu2022doped} have been reported by hole  doping of the CSLs based on 
  density matrix renormalization group (DMRG) studies of ladder systems. Meanwhile, numerical simulations have identified $d+id$-wave SC besides  topologically trivial nematic $d$-wave SC after doping the CSL or nearby magnetically ordered phases on a triangular-lattice spin system with  TRS-breaking three-spin chiral interactions~\cite{jiang2020topological, huang2021topological}.  Interestingly, $d+id$-wave SC also appears when doping the antiferromagnetic phase of the triangular-lattice $J_1-J_2$ model through spontaneous TRS breaking~\cite{huangQuantumPhaseDiagram2023c} in a narrow parameter window. These observations demonstrate the competing orders in doped CSL and the rich interplay between geometrical frustration, hole dynamics, and spin fluctuations~\cite{fradkin2015colloquium,song2021doping}.

Experimentally,   candidate materials for spin liquid are rare~\cite{khatuaExperimentalSignaturesQuantum2023,knolleFieldGuideSpin2019}, and the route from spin liquid to unconventional SC remains illusive. 
Recently, the twisted transition metal dichalcogenide moiré system has emerged as a highly tunable platform for exploring strong correlation effects on a triangular lattice~\cite{wu2018hubbard,Tang2020}. Particularly, the large moiré lattice constants make a large magnetic flux per unit cell accessible with currently available laboratory fields. Under a magnetic field, this system can be described by the Hubbard-Hofstadter model with spin or pseudo-spin SU(2) symmetry~\cite{kuhlenkampChiralPseudospinLiquids2024b}, which features strong correlation, nontrivial band topology, TRS breaking, and frustration. At half-filling, a CSL phase has been found between the integer quantum Hall (IQH) phase at weak coupling and the 120\textdegree{} Néel ordered phase at strong coupling limit~\cite{kuhlenkampChiralPseudospinLiquids2024b,divic2024chiralspinliquidquantum}. Electron pairing has been numerically confirmed near the half-filling over a wide range of coupling strength on both sides of the IQH-CSL critical point~\cite{divic2024anyonsuperconductivitytopologicalcriticality}.  It is argued through parton mean-field theory that topological SC emerges after doping near the topological quantum criticality associated to the closing of the charge gap while the spin gap remains open~\cite{divic2024anyonsuperconductivitytopologicalcriticality} (this argument may be generalized to fractional fillings~\cite{pichler2025microscopicmechanismanyonsuperconductivity}).  This naturally leads to open questions such as whether the chiral SC survives competing orders at finite doping and what its topological nature is 
 from doping either the IQH or CSL phases.  Answering these questions will constitute a significant step towards  establishing the mechanism of anyon superconductivity near topological criticality in realistic strongly correlated systems.

\begin{figure}
   \includegraphics[width=0.48\textwidth,angle=0]{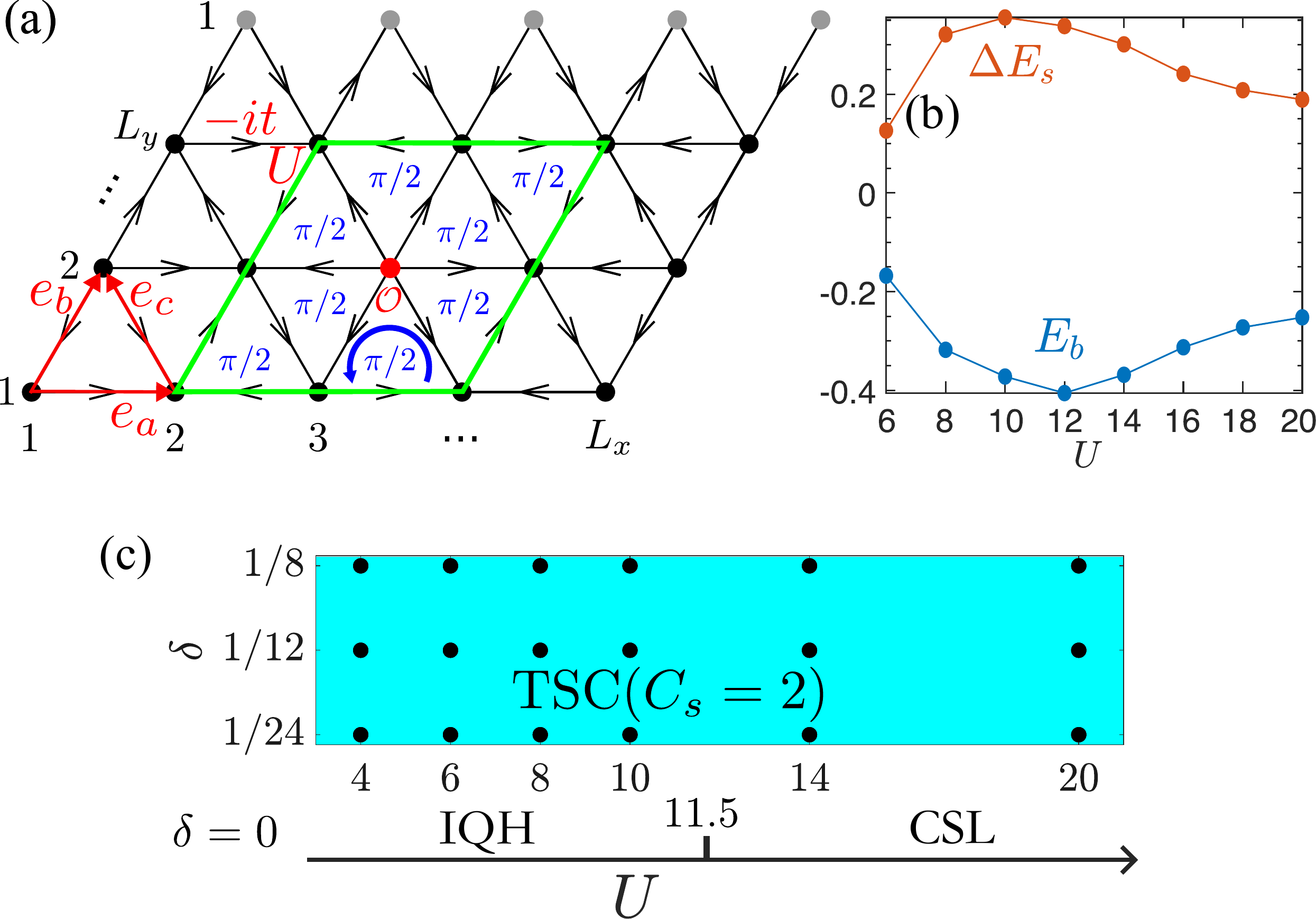}
   \caption{\label{Latt} (a) The Hofstadter Hubbard model on a triangular lattice. Each triangle is inserted $\pi/2$ flux for both spins. The arrow denotes the positive direction of $\tau_{ij}$ in Eq.~\ref{Hm}. The green rectangular denotes the unit cell. (b) Cooper pair binding energy $E_b=E(N_e+2,0)+E(N_e,0)-2E(N_e+1,1/2)$ and spin excitation gap $\Delta E_s=E(N_e,1)-E(N_e,0)$ at different $U$s, obtained at $N=36\times 6$ and $\delta=1/12$. Here $E(n,S_z^\text{tot})$ is the ground-state energy for the electron number $n$ and total spin $S_z^\text{tot}$. (c) Phase diagram of the doped 6-leg Hubbard-Hofstadter model. Black dots represent the parameter points where DMRG calculations are performed. A topological SC (TSC) phase with spin Chern number $C_s=2$ emerges after doping both IQH and CSL phases.
   A half-filling phase diagram is also shown for comparison. }
\end{figure}

In this work, we address these questions through two unbiased and controlled numerical simulation methods.
Through large-scale DMRG simulations~\cite{white1992density} of the lightly-doped triangular-lattice Hubbard-Hofstadter model with one-quarter flux quantum per triangle plaquette, we observe  chiral SC over a broad range of the coupling strength $U$ (Fig.~\ref{Latt}(c)), corresponding to parent states ranging from IQH to CSL phases. The chiral SC is characterized by quasi-long-range pairing correlations, and a quantized spin Chern number $\mathcal{C}_s=2$ with possibly coexisting subdominant CDW. 
From a finite-temperature perspective, we apply determinantal quantum Monte Carlo algorithm (DQMC)~\cite{DQMC1,DQMC2} to this model and find that SC pairing fluctuations become increasingly dominant as the temperature is lowered.

The Hamiltonian of the triangular-lattice Hubbard-Hofstadter model has the form
\begin{equation}
H=-it\sum_{\sigma}\sum_{\langle xy\rangle} \tau_{ij}(c_{i\sigma}^{\dagger} c_{j\sigma}- c_{j\sigma}^{\dagger} c_{i\sigma})+ U\sum_{i} n_{i\uparrow}n_{i\downarrow},\notag \\
\label{Hm}
\end{equation}
where $\sigma=\pm$ labels spins and $\tau_{ij}=+1$ or -1 if $j$ to $i$ follows or is opposite to the arrows in Fig.~\ref{Latt}. This chosen gauge is called imaginary $C_{6z}$ gauge 
and the Hamiltonian is symmetric under $\pi/3$ bare rotation around $\mathcal{O}$ in Fig.~\ref{Latt}.  We choose $t=1$ as the energy unit. The hole doping level is denoted by $\delta$, rendering the total electron number $N_e=N(1-\delta)$. For DMRG, we consider a triangular lattice on a cylinder with $N=L_x \times L_y$ sites, and impose the periodic (open) boundary condition in the $y$ ($x$) direction (see Fig.~\ref{Latt}(a)). In the main text, we focus on $L_y=6$ and present additional results for $L_y=3,4,8$ in the Supplemental Material~\cite{SuppMaterial}. The simulations enforce both charge-conservation U(1) and spin-SU(2) symmetries~\cite{McCulloch2007}, and retain up to $M = 16000$ SU(2) Schmidt states (bond dimension) for large $L_y=6$ systems, resulting in a truncation error $\epsilon \approx 5.0 \times 10^{-6}$.
For DQMC, we simulate the triangular lattice with periodic boundary conditions along both directions (torus geometry), using a $6 \times 6$ system size in the main text, with results for the smaller $4\times4$ and larger $8\times8$ systems 
system discussed and compared in the Supplemental Material~\cite{SuppMaterial},
showing qualitatively consistent  conclusions.

\begin{figure}
   \includegraphics[width=0.48\textwidth,angle=0]{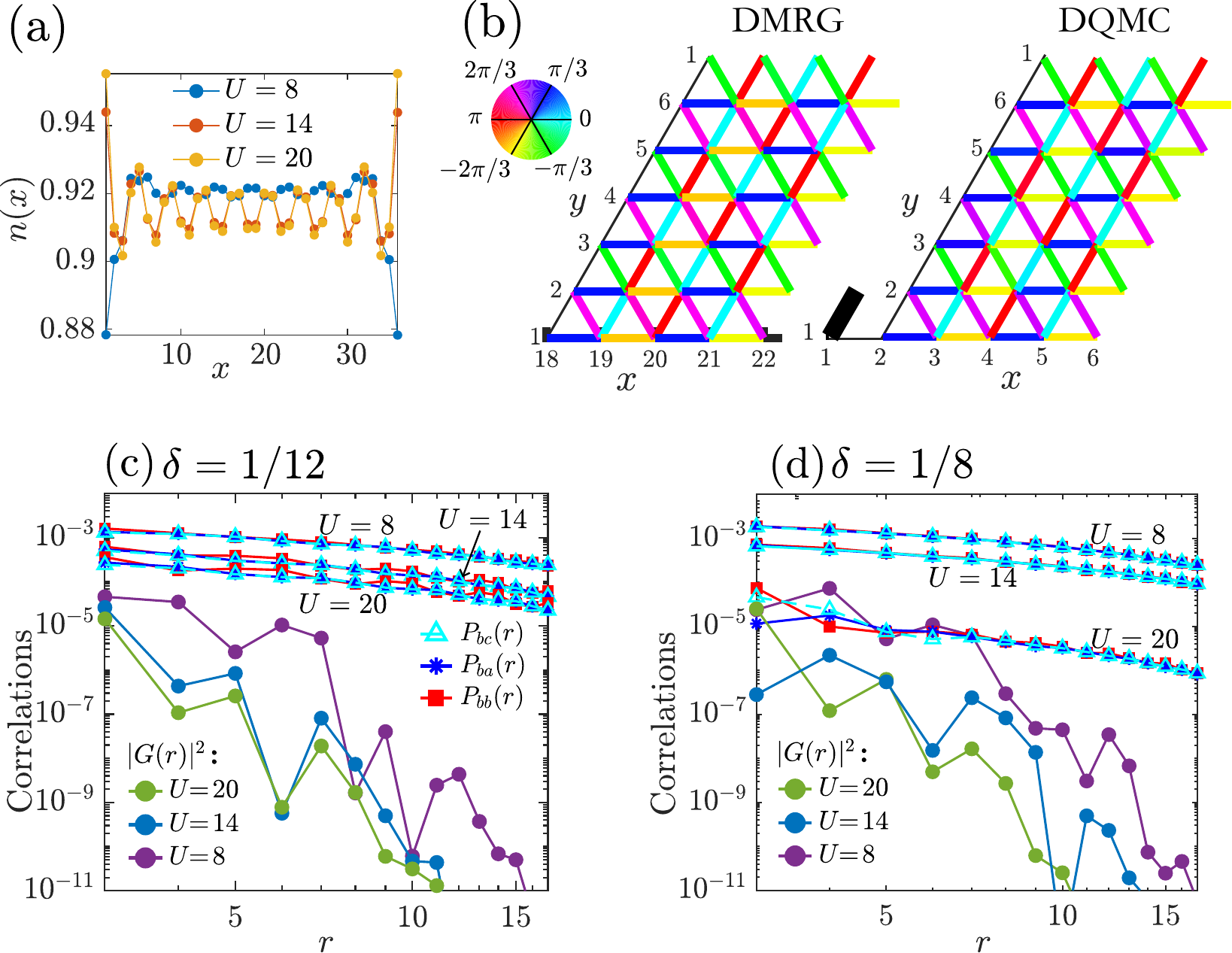}
   \caption{\label{Occu_corr} (a) Rung-averaged electron density profiles $n(x)=\sum_{y=1}^{L_y}(n_{(x,y),\uparrow}+n_{(x,y),\downarrow})/L_y$ for $N=36\times 6$ at various $U$s and $\delta=1/12$. (b) The phase pattern of pairing on different bonds in the chiral SC phase represented by $U=14$ and $\delta=1/12$. 
   The phase of the SC order on a bond $\alpha$ at $\mathbf{r}$ with respect to that of the reference bond along $\mathbf{e_b}$ at $\mathbf{r_0}=(13,1)$: $\text{arg}(\langle \hat{\Delta}^\dagger_b(\mathbf{r}_0)\hat{\Delta}_\alpha(\mathbf{r})\rangle)=\text{arg}(\langle\hat{\Delta}_\alpha(\mathbf{r})\rangle)-\text{arg}(\langle\hat{\Delta}_b(\mathbf{r_0})\rangle)$ are shown. DQMC data are simulated on a $6\times 6$ lattice with $U/t=8$, $\delta=0.02$, and $T/t=1/4$, using correlations at imaginary time $\tau=\beta/2$. (c) Pair-Pair correlations $P_{\alpha\beta}(r)$ and single-particle Green's function $G(r)=\sum_\sigma\langle \hat{c}^\dagger_\sigma(x_0,y_0) \hat{c}_\sigma(x_0+r,y_0)$ for various $U$s and $\delta=1/12$ and $N=36\times 6$. (d) A similar plot for $\delta=1/8$ and $N=32\times 6$.}
\end{figure}

{\it Quantum phase diagram--} To systematically identify the phase diagram across a range of doping and interaction strengths, we first perform a comprehensive set of DMRG simulations.  In Fig.~\ref{Occu_corr}(a),  we show the electron density profiles  corresponding to doping  the  IQH phase at  $U=8$ and the CSL phase near ($U=14$) and far from ($U=20$) criticality, respectively.  In all three cases, there are charge stripes with a period of 4 sites along the $x$ direction, corresponding to two holes per stripe, but the CDW amplitude is much weaker in the doped IQH phase. The phase structure of the nearest-neighbor pairing in Fig.~\ref{Occu_corr}(b) demonstrates that SC order is antisymmetric under magnetic translations along both $e_a$ and $e_b$~\cite{SuppMaterial}, corresponding to one of the four one-dimensional irreducible representations of the magnetic translational group when the flux per unit cell is $\pi$~\cite{shafferUnconventionalSelfsimilarHofstadter2022,shafferTheoryHofstadterSuperconductors2021}. The pairing order gains $-\pi/3$ phase under the $C_6$ magnetic rotation, which takes its bare form when the center of rotation is $\mathcal{O}$. 
In Fig.~\ref{Occu_corr}(c-d), we show the dominant spin singlet pair-pair correlations $P_{\alpha\beta}(r)=\langle  \hat{\Delta}^\dagger_\alpha(x_0,y_0)\hat{\Delta}_\beta(x_0+r,y_0)\rangle$ with the pairing order $\Delta_\alpha(x,y)
=(\hat{c}_{(x,y),\uparrow}\hat{c}_{(x,y)+\mathbf{e}_{\alpha},\downarrow}-\hat{c}_{(x,y),\downarrow}\hat{c}_{(x,y)+\mathbf{e}_{\alpha},\uparrow} )/\sqrt{2}$, where  $\alpha = a,b,c$ and $(x,y)$ denotes the location $x\mathbf{e}_a+y\mathbf{e}_b$. 
For both doping levels $\delta=1/12$ and $1/8$, the pairing correlations are strong and decay slowly with distances for $U$ from $8$ to  $14$, but they are suppressed by one order of magnitude for $U=20$ at $\delta=1/8$, indicating weakened SC for very strong $U$.   The isotropy of the chiral SC order is demonstrated by $P_{ba}\approx P_{bb}\approx P_{bc}$. The pair-pair correlations are much stronger than the amplitude squared of the single-particle Green's function $|G(r)|^2$ at longer distances, demonstrating the charge-2$e$ mode as the  origin of SC. 
Consistently, the electrons show strong binding ($E_b<0$) while the spin gap is finite ($\Delta E_s>0$)
 (Fig.~\ref{Latt}(b)). 
 The corresponding phase diagram is presented in Fig.~\ref{Latt}(c)
with the chiral SC phase present for a wide range of $U$ and doping levels.

On the other hand, for smaller $U$ (i.e., $U\lesssim3$), SC becomes too weak to be distinguished from a possible metallic phase, as indicated by  the absence of clearly dominant correlations (see Fig. S11~\cite{SuppMaterial}), though a perturbative analysis suggests that an instability toward superconductivity may still emerge in the weak-coupling limit~\cite{SuppMaterial}. Taken together, the suppression of SC at both small and large $U$ indicates that SC is more robust at intermediate interaction strength, approximately near the topological criticality of the undoped parent system~\cite{divic2024anyonsuperconductivitytopologicalcriticality}.

\begin{figure}
   \includegraphics[width=0.48\textwidth,angle=0]{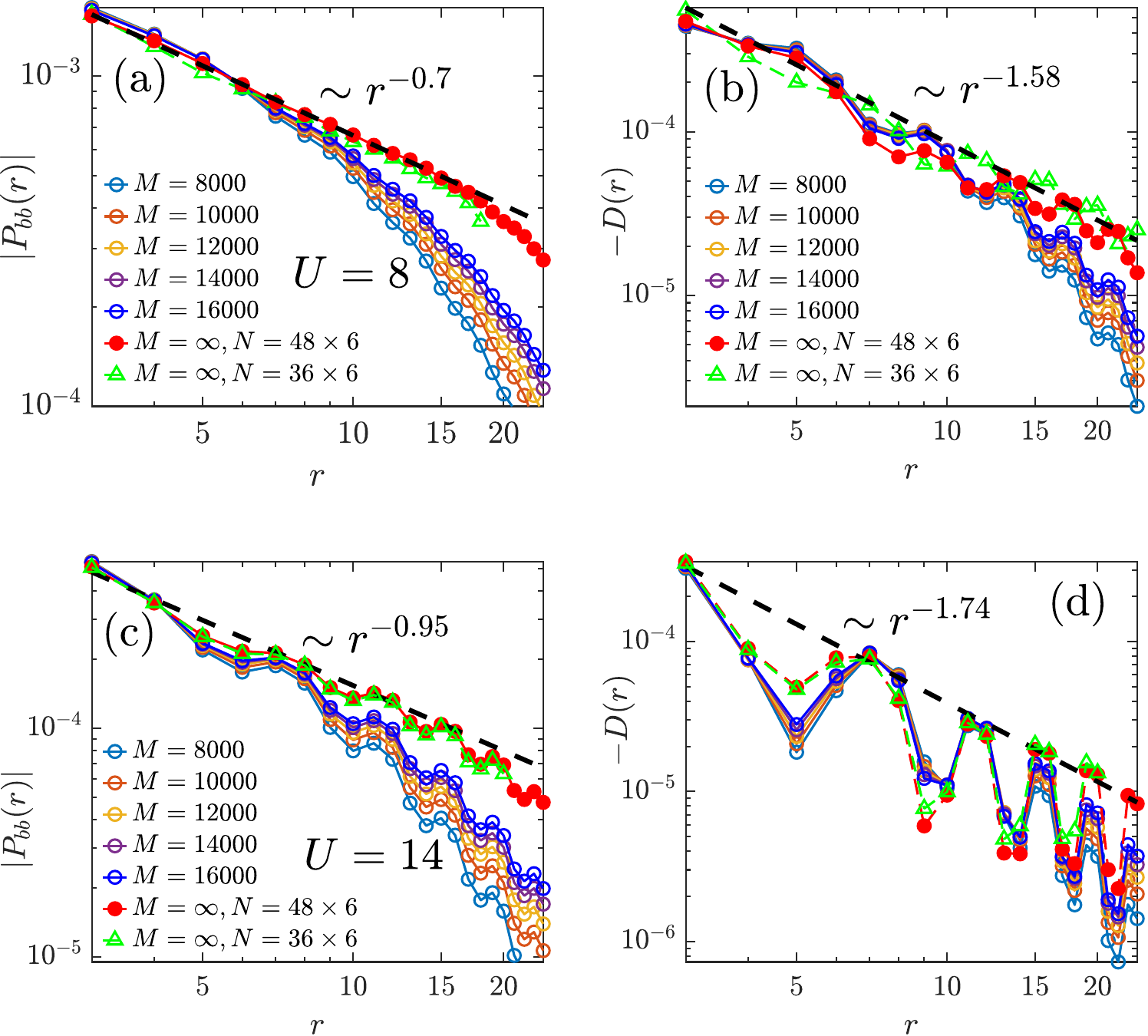}
   \caption{\label{corr} (a-b) The scalings and power-law fittings of pair-pair correlation $P_{bb}(r)$ and density-density correlation $D(r)$ for $U=8$ and $\delta=1/12$. (c-d) Similar plots for $U=14$. The second-order extrapolation to infinite $M$ is applied. The results for $N=48\times 6$ and $N=36\times 6$ are consistent.}
\end{figure}

{\it Dominant pairing correlations--} Two representative points corresponding to doping the IQH ($U=8$) and CSL ($U=14$) states are characterized in Fig.~\ref{corr} for systems size $N=48\times 6$ at doping level $\delta=1/12$.
The magnitude of pairing correlations at longer distances $\left | P_{bb}(r) \right |$  increases gradually as the DMRG bond dimension increases from $M=8000$ to $16000$.
Because the DMRG represents the ground state in the matrix product form~\cite{schollwock2011density} with finite bond dimensions, the scaling to $M\rightarrow \infty$ is needed to identify the true nature of long-distance correlations for wide cylinders. 
Using a second-order polynomial fitting of $1/M$, we find that the extrapolated $\left | P_{bb}(r) \right |$ shows a power-law decay with distance  $\left | P_{bb}(r) \right |\sim r^{-K_\text{SC}}$, with the Luttinger exponent $K_\text{SC}\approx 0.70$, and $0.95$ for $U=8$ and $14$, respectively. Similar results are obtained for correlations with other bonds and for a smaller
size $N=36\times 6$.  $K_\text{SC}\lesssim 1$  indicates strong divergent SC susceptibility in the zero-temperature limit when $L_x\rightarrow \infty$~\cite{jiang2021superconductivity}. Similarly, quasi-long-range CDW orders
are observed through the density-density correlations $D(r) = \langle {\hat{n} }_{x_0,y_0} {\hat{n} }_{x_0+r,y_0} \rangle - \langle {\hat{n} }_{x_0,y_0} \rangle \langle {\hat{n} }_{x_0+r,y_0} \rangle\sim r^{-K_\text{CDW}}$ with a larger Luttinger exponent
($K_\text{CDW}\approx 1.58$ and $1.74$ for $U=8$ and $14$, respectively), indicating the dominance of SC order. The Luttinger exponent for SC decreases with the cylinder width~\cite{SuppMaterial}, suggesting its presence in the two-dimensional limit.

\begin{figure}
   \includegraphics[width=0.48\textwidth,angle=0]{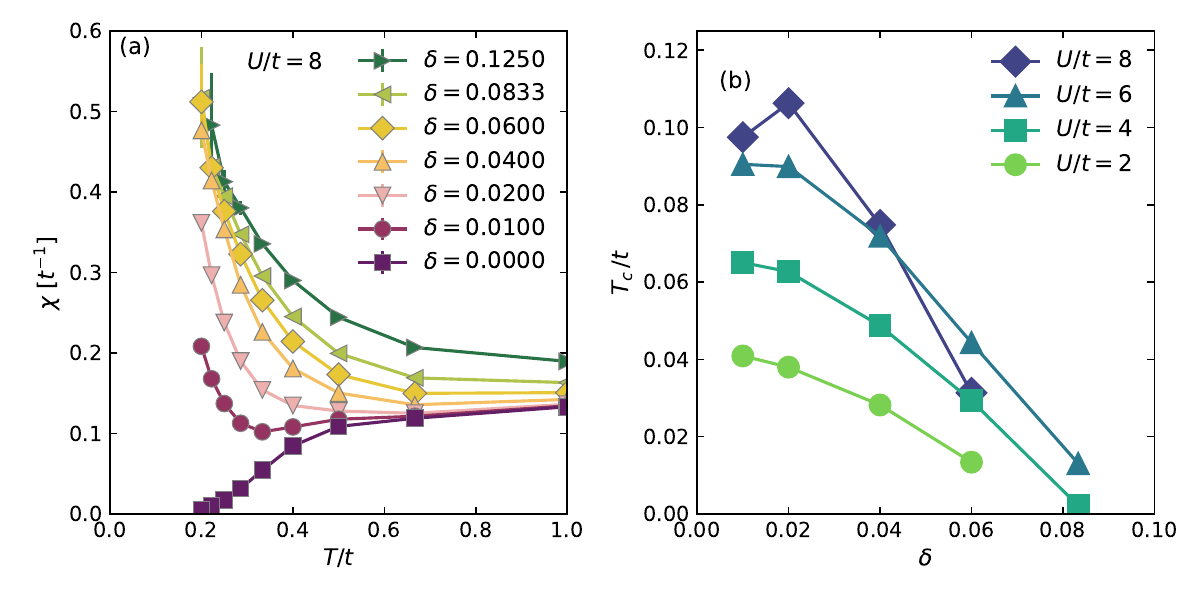}
   \caption{\label{Tcvp}
(a) Temperature dependence of pairing susceptibility $\chi(\tau=\beta/2)$ for $U/t = 8$.  
   (b) Superconducting characteristic temperature $T_c$ 
   estimated by linearly extrapolating $\chi^{-1}(\tau=\beta/2)$ to zero.
   Simulation lattice size is $6\times 6$.}
\end{figure}

As a complement to the DMRG simulations, which detect the orders of ground state through equal-time correlation functions, the unequal-time correlation function  $\chi(\beta / 2)\equiv \beta\langle\hat{O}^{\dagger}(\tau=\beta/2) \hat{O}(\tau=0)\rangle/N$ measured in DQMC provides a dynamic perspective on low-temperature order by probing the corresponding low-energy bosonic fluctuations within an energy window of order $k_B T$~\cite{SuppMaterial}. Specifically, as shown in the right panel of Fig.~\ref{Occu_corr}(b), the pairing phase pattern at temperature $T = t/4$ for intermediate interaction strength $U = 8$ can be extracted by $\langle \Delta^{\dagger}_\alpha(x_0, y_0,\tau=\beta/2) \Delta_\beta(x_0+r, y_0,\tau=0)\rangle$, and is in excellent agreement with that identified by DMRG. 
At the same temperature, the pattern becomes disrupted at significantly smaller $U$ or at half-filling, as shown in the Supplementary Material\cite{SuppMaterial} (Fig.~\ref{dqmc_phase_comparison}).  
Once the pairing phase pattern is determined, the unequal-time pairing susceptibility $\chi(\beta/2)$ can be defined by taking $\hat{O} = \sum_{\alpha,x,y} \Delta_\alpha(x, y) e^{-i\theta_\alpha(x, y)}$, where the sum runs over all inequivalent bonds $(x, y, \alpha)$ with their corresponding phase factors $\theta_{\alpha}(x, y)$.
As illustrated in Fig.~\ref{Tcvp}(a), $\chi(\beta / 2)$ increases rapidly with decreasing temperature at finite doping $\delta \neq 0$, indicating that pairing fluctuations are increasingly transferred into the low-energy window bounded by the thermal scale $k_B T$. In contrast, for $\delta = 0$, this enhancement is absent, suggesting that pairing fluctuations remain irrelevant in the undoped case, as theoretically expected. Upon extrapolating the data to lower temperatures, we define a characteristic temperature $T_c$ as the point where $\chi(\beta / 2)$ diverges, signaling the proliferation of low-energy pairing modes (c.f. Supplementary Material\cite{SuppMaterial} (Fig.~\ref{extrapolation})). This $T_c$ serves as an alternative temperature scale for SC, physically analogous to the standard critical temperature defined by the divergence of the static (zero-frequency) susceptibility. The dependence of the estimated $T_c$ on doping and interaction strength is summarized in Fig.~\ref{Tcvp}(b). Notably, the dome-like doping dependence of $T_c$, which vanishes at zero doping, suggests that SC fluctuations are intimately connected to the nature of the undoped state. Moreover, in the DQMC-accessible range $U \leq 8$, which already approaches the intermediate-coupling regime near the DMRG-identified critical point at half-filling~\cite{divic2024chiralspinliquidquantum}, $T_c$ increases with $U$, indicating enhanced pairing correlations.

\begin{figure}
   \includegraphics[width=0.48\textwidth,angle=0]{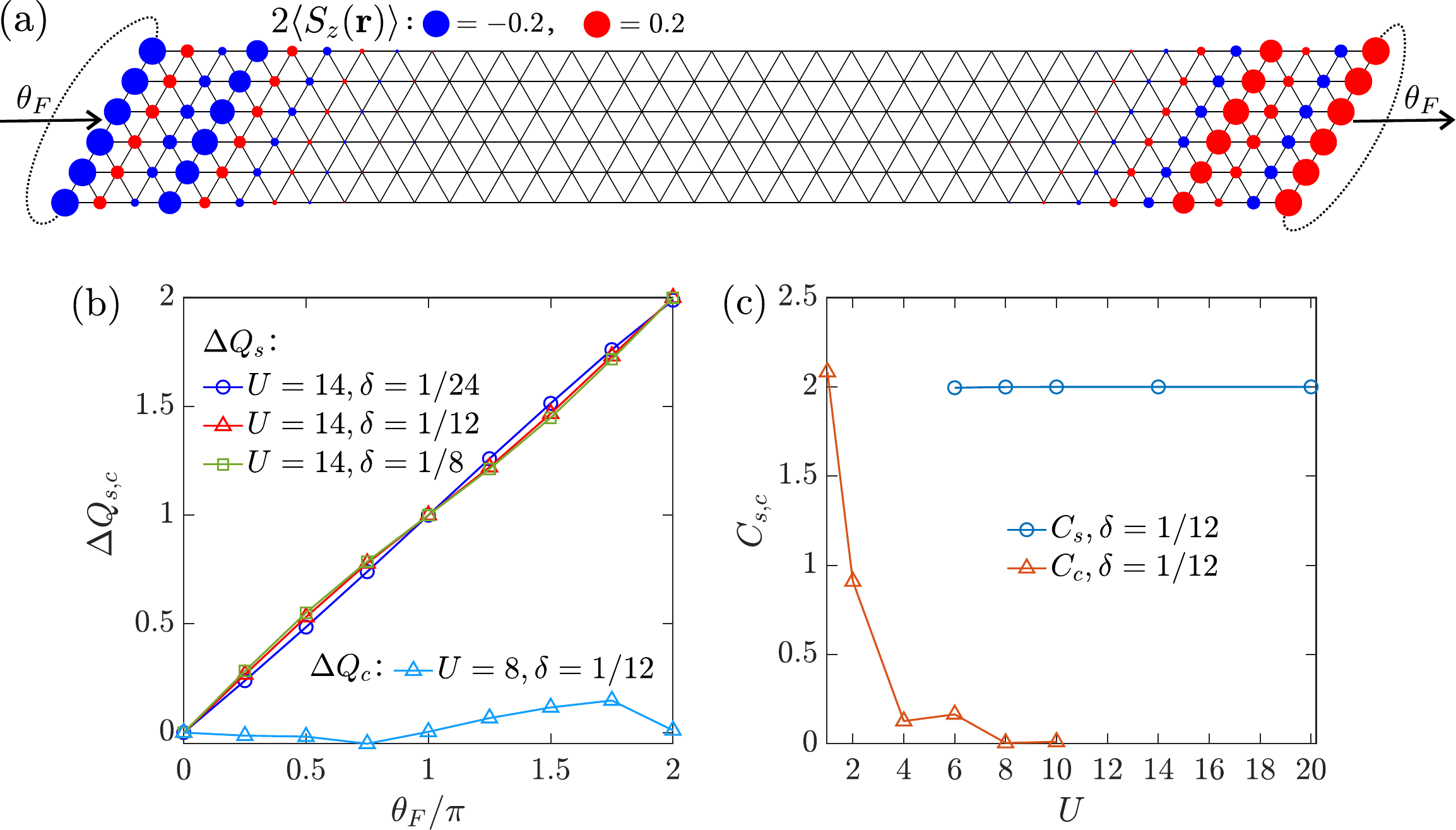}
   \caption{\label{chern} (a) Real-space configuration of the spin magnetization after adiabatically inserting a quantized flux $\theta_F$. The area of the circle is proportional to the amplitude of 2$\langle S_z(\mathbf{r})\rangle$. The red (blue) color represents the positive (negative) value. (b) The net spin (charge) transferred $\Delta Q_{s,c}$ from the left edge to the right one after adiabatically threading spin-dependent (-independent) flux $\theta_F$ with steps of $\pi/4$. (c) Spin and charge Chern numbers $C_{s,c}=\Delta Q_{s,c}(\theta_F=0\rightarrow 2\pi)$ at different $U$ and $\delta$. The obtained results are well converged with a U(1) bond dimension $m=10000$. }
\end{figure}

{\it Quantized spin Chern number--}To characterize the topological nature of the TSC state, we perform the flux insertion through the cylinder  in the $U(1)\times U(1)$ DMRG simulations (Fig.~\ref{chern}(a))~\cite{gong2014,zaletel2014flux}. Specifically,  we impose the spin-dependent boundary condition  $\hat{c}_{(x,L_y+1),\sigma}=\hat{c}_{(x,1),\sigma}e^{i\sigma\theta_F}$ and measure the accumulated spins $\Delta Q_s=(n_\uparrow-n_\downarrow)|^{\theta_F}_0$  near the right edge while adiabatically increasing $\theta_F$ from 0 to $2\pi$ (Fig.~\ref{chern}(b)).  Similarly, a spin-independent boundary condition $\hat{c}_{(x,L_y+1),\sigma}=\hat{c}_{(x,1),\sigma}e^{i\theta_F}$ is used to calculate the accumulated charges $\Delta Q_c=(n_\uparrow+n_\downarrow)|^{\theta_F}_0$ near the right edge. 
The spin Chern number is given by $C_s=\Delta Q_s(\theta_F=0\rightarrow 2\pi)$ and has a well-quantized value of 2  in the TSC phase (Fig.~\ref{chern}(c)), similar to other TSC states in doped triangular Mott insulators without considering the orbital effect of complex hopping~\cite{huangQuantumPhaseDiagram2023c,huang2021topological}. In contrast, the charge Chern number $C_c=\Delta Q_c(\theta_F=0\rightarrow 2\pi)$ vanishes (or is nearly zero) in the TSC phase, but remains finite in the smaller $U$ regime ($U\lesssim2$) possibly consistent with chiral metal behavior in our finite size systems. The TSC state is further characterized by the nonzero spin chiral order $\left \langle \chi \right \rangle = \langle \hat{\boldsymbol{S}}_{i}\cdot (\hat{\boldsymbol{S}}_{j}\times \hat{\boldsymbol{S}}_{k}) \rangle$ ($i,j,k$ are the sites of the elementary triangle plaquettes) (see Supplementary for details~\cite{SuppMaterial}).

\begin{figure}
   \includegraphics[width=0.48\textwidth,angle=0]{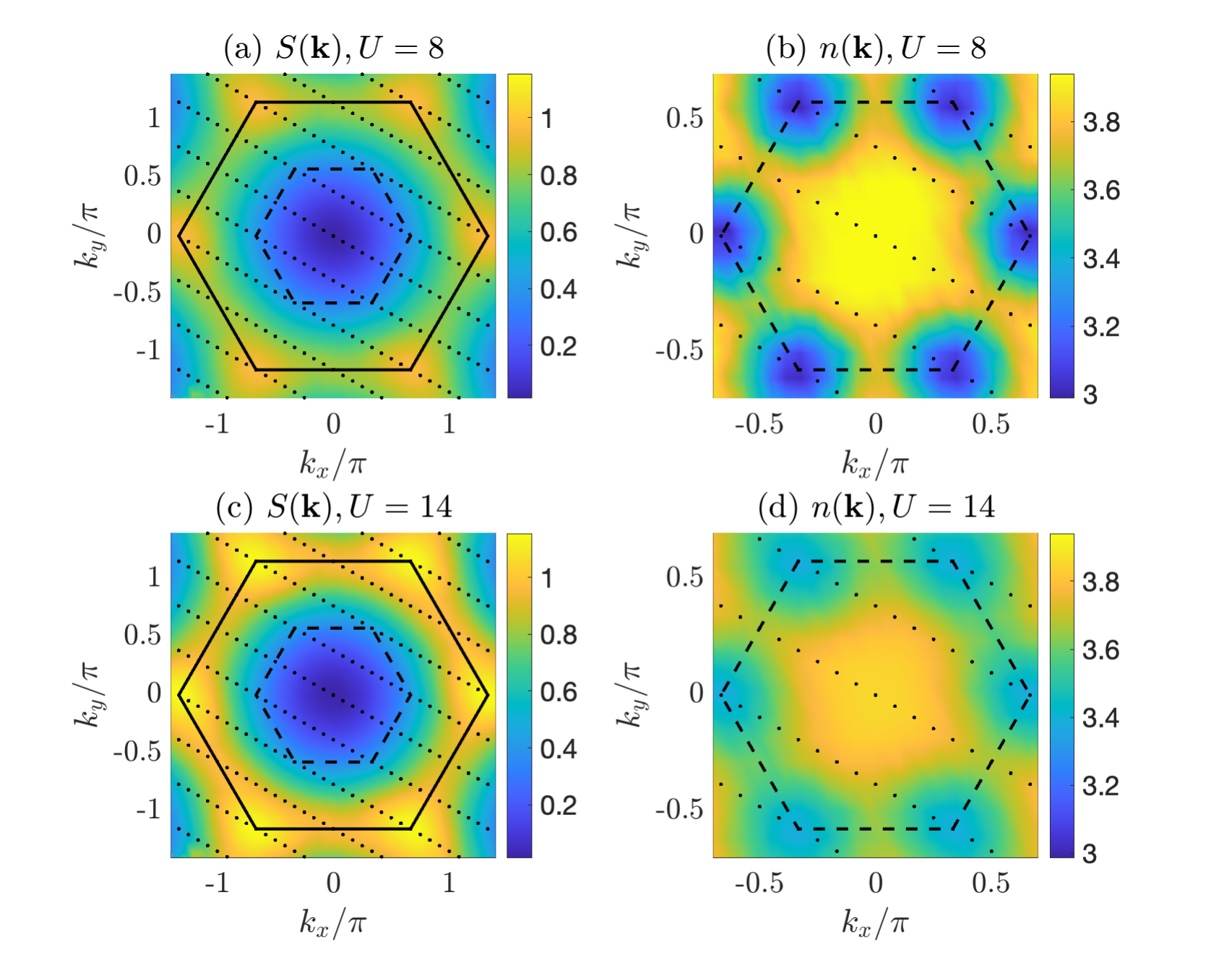}
   \caption{\label{Struct_factor}(a,c) Spin structure factors  $S(\mathbf{k})$ and (b,d) momentum-space electron density distribution $n(\mathbf{k})$ for $U=8$ and 14, respectively. The first magnetic Brillouin zone is shown by the dashed black line, and the first non-magnetic Brillouin zone by the solid black line. The accessible wavevectors  are marked by black dots. Parameters:  $N=48\times 6$ and $\delta=1/12$.}
\end{figure}

{\it Summary and Discussion--} To summarize, through ground-state DMRG and finite-temperature DQMC simulations of the triangular Hofstadter-Hubbard model, we identify a single chiral TSC phase with spin Chern number 2 and long-range spin chiral order across a wide range of interaction strength corresponding to doping both the IQH and CSL phases.  
Our results thus support the possibility of experimentally realizing TSC in the twisted transition metal dichalcogenide Moiré systems. 
The strongest electron binding is observed near the topological critical point between IQH and CSL. The quasi-long-range TSC is possibly coexisting with subdominant  CDW in the  quasi-1D systems, both of which grow more robust with the system width\cite{SuppMaterial}.  The SC order gains $-\pi/3$ phase under the magnetic $C_6$ rotation and is antisymmetric under unit magnetic translations.

In the future, it would be particularly interesting to study the normal state properties in more detail, in order to distinguish between BCS-like and anyonic pairing mechanisms.  Our study demonstrates a single zero temperature TSC phase over a wide range of interaction strengths [Fig.~\ref{Latt}(c)], which may be attributed to topological criticality, which coherently softens a charge‑2e bosonic mode and thereby energetically stitches the two sides together~\cite{divic2024anyonsuperconductivitytopologicalcriticality}.  However, the normal states for $T>T_c$ may be quite distinct in different interaction ranges.  In the small-$U$ regime, the doped system has a well-defined Fermi surface [Fig.~\ref{Struct_factor}(a,b) and Fig.~\ref{Supp_spinSf} in Supplementary Material~\cite{SuppMaterial}], and 
second-order perturbative processes via inter-band transitions~\cite{SuppMaterial, PhysRevB.105.094506} 
provide a natural pairing route --- basically similar to the Fermi surface instability induced by attractive channels in conventional metals. Therefore, an essentially Fermi‑liquid metal with a non‑quantized Hall response
~\cite{PhysRevB.100.115102, MIKou} is expected for $T>T_c$.  On the other hand, as $U$ increases, the Fermi surface gradually disappears [Fig.~\ref{Struct_factor}(c,d)], electronic correlations become dominant, and the TSC appears to emerge directly from a doped CSL. This crossover is supported by the evolution of the ratio between the bare (bubble) susceptibility—constructed from single-particle Green’s functions in the DQMC simulations—and the fully renormalized susceptibility, as shown in the Supplementary Material\cite{SuppMaterial} (Fig.~\ref{ratio}). While this ratio remains close to unity at weak coupling, it is strongly suppressed at large $U$, extrapolating to zero even at finite temperature, signaling that the pairing channel is driven by strong correlation effects.   Future studies should probe the view of this normal state as a compressible anyon fluid~\cite{shi2024dopingfractionalquantumanomalous, PhysRevB.111.014508, shi2025anyondelocalizationtransitionsdisordered}, with an eye to experimentally measurable signatures.  Our results may be a starting point for this endeavor.

\textit{Note Added.---} At the stage of finalizing our work, we learned of a parallel study\cite{preprint} using infinite DMRG, which obtained results consistent with ours.

\textit{Acknowledgments.---} D.N.S. acknowledges  helpful discussions with Clemens Kuhlenkamp and Ashvin Vishwanath. W.O.W. acknowledges support from the Gordon and Betty Moore Foundation through Grant GBMF8690 to the University of California, Santa Barbara, to the Kavli Institute for Theoretical Physics (KITP).
This research was supported in part by grant NSF PHY-2309135 to the KITP.
DQMC simulations made use of computational facilities purchased with funds from the NSF (CNS-1725797) and administered by the Center for Scientific Computing (CSC). The C.S.C. is supported by the California NanoSystems Institute and the Materials Research Science and Engineering Center (MRSEC; NSF DMR 2308708) at UC Santa Barbara. J.~X.~Z. was supported by the European Research Council (ERC) under the European Union’s Horizon 2020 research and innovation program (Grant Agreement No.\ 853116, acronym TRANSPORT). L.B. acknowledges support by the DOE through grant No. DE-SC0020305 for theoretical analysis of superconductivity, and the Simons Collaboration on Ultra-Quantum Matter, which is a grant from the Simons Foundation (Grant No. 651440) for guidance of numerics. D.N.S. was  supported by  the DOE, Office of BES under Grant No. DE-FG02-06ER46305 for large scale  simulations of interacting systems. We also acknowledge NSF grant No.\ DMR-2406524 (D.N.S.) for instrument support. 

\textit{Data availability.---}
Data and simulation codes are available from authors upon reasonable request.

\textit{Author contributions.---}
F.C. performed the DMRG studies.
W.O.W. conceptualized and performed the DQMC studies.
J.-X.Z. carried out the analytical calculations and contributed to the interpretation of the DQMC results.
All authors contributed to the development and completion of this work. 

\newpage
\bibliography{Tri}

\begin{thebibliography}{54}%
\makeatletter
\providecommand \@ifxundefined [1]{%
 \@ifx{#1\undefined}
}%
\providecommand \@ifnum [1]{%
 \ifnum #1\expandafter \@firstoftwo
 \else \expandafter \@secondoftwo
 \fi
}%
\providecommand \@ifx [1]{%
 \ifx #1\expandafter \@firstoftwo
 \else \expandafter \@secondoftwo
 \fi
}%
\providecommand \natexlab [1]{#1}%
\providecommand \enquote  [1]{``#1''}%
\providecommand \bibnamefont  [1]{#1}%
\providecommand \bibfnamefont [1]{#1}%
\providecommand \citenamefont [1]{#1}%
\providecommand \href@noop [0]{\@secondoftwo}%
\providecommand \href [0]{\begingroup \@sanitize@url \@href}%
\providecommand \@href[1]{\@@startlink{#1}\@@href}%
\providecommand \@@href[1]{\endgroup#1\@@endlink}%
\providecommand \@sanitize@url [0]{\catcode `\\12\catcode `\$12\catcode `\&12\catcode `\#12\catcode `\^12\catcode `\_12\catcode `\%12\relax}%
\providecommand \@@startlink[1]{}%
\providecommand \@@endlink[0]{}%
\providecommand \url  [0]{\begingroup\@sanitize@url \@url }%
\providecommand \@url [1]{\endgroup\@href {#1}{\urlprefix }}%
\providecommand \urlprefix  [0]{URL }%
\providecommand \Eprint [0]{\href }%
\providecommand \doibase [0]{https://doi.org/}%
\providecommand \selectlanguage [0]{\@gobble}%
\providecommand \bibinfo  [0]{\@secondoftwo}%
\providecommand \bibfield  [0]{\@secondoftwo}%
\providecommand \translation [1]{[#1]}%
\providecommand \BibitemOpen [0]{}%
\providecommand \bibitemStop [0]{}%
\providecommand \bibitemNoStop [0]{.\EOS\space}%
\providecommand \EOS [0]{\spacefactor3000\relax}%
\providecommand \BibitemShut  [1]{\csname bibitem#1\endcsname}%
\let\auto@bib@innerbib\@empty
\bibitem [{\citenamefont {Keimer}\ \emph {et~al.}(2015)\citenamefont {Keimer}, \citenamefont {Kivelson}, \citenamefont {Norman}, \citenamefont {Uchida},\ and\ \citenamefont {Zaanen}}]{keimer2015quantum}%
  \BibitemOpen
  \bibfield  {author} {\bibinfo {author} {\bibfnamefont {B.}~\bibnamefont {Keimer}}, \bibinfo {author} {\bibfnamefont {S.~A.}\ \bibnamefont {Kivelson}}, \bibinfo {author} {\bibfnamefont {M.~R.}\ \bibnamefont {Norman}}, \bibinfo {author} {\bibfnamefont {S.}~\bibnamefont {Uchida}},\ and\ \bibinfo {author} {\bibfnamefont {J.}~\bibnamefont {Zaanen}},\ }\bibfield  {title} {\bibinfo {title} {From quantum matter to high-temperature superconductivity in copper oxides},\ }\href {https://www.nature.com/articles/nature14165} {\bibfield  {journal} {\bibinfo  {journal} {Nature}\ }\textbf {\bibinfo {volume} {518}},\ \bibinfo {pages} {179} (\bibinfo {year} {2015})}\BibitemShut {NoStop}%
\bibitem [{\citenamefont {Lee}\ \emph {et~al.}(2006)\citenamefont {Lee}, \citenamefont {Nagaosa},\ and\ \citenamefont {Wen}}]{lee2006doping}%
  \BibitemOpen
  \bibfield  {author} {\bibinfo {author} {\bibfnamefont {P.~A.}\ \bibnamefont {Lee}}, \bibinfo {author} {\bibfnamefont {N.}~\bibnamefont {Nagaosa}},\ and\ \bibinfo {author} {\bibfnamefont {X.-G.}\ \bibnamefont {Wen}},\ }\bibfield  {title} {\bibinfo {title} {Doping a mott insulator: Physics of high-temperature superconductivity},\ }\href {https://doi.org/10.1103/RevModPhys.78.17} {\bibfield  {journal} {\bibinfo  {journal} {Rev. Mod. Phys.}\ }\textbf {\bibinfo {volume} {78}},\ \bibinfo {pages} {17} (\bibinfo {year} {2006})}\BibitemShut {NoStop}%
\bibitem [{\citenamefont {Anderson}(1987)}]{anderson1987resonating}%
  \BibitemOpen
  \bibfield  {author} {\bibinfo {author} {\bibfnamefont {P.~W.}\ \bibnamefont {Anderson}},\ }\bibfield  {title} {\bibinfo {title} {The resonating valence bond state in la2cuo4 and superconductivity},\ }\href {https://www.science.org/doi/10.1126/science.235.4793.1196} {\bibfield  {journal} {\bibinfo  {journal} {Science}\ }\textbf {\bibinfo {volume} {235}},\ \bibinfo {pages} {1196} (\bibinfo {year} {1987})}\BibitemShut {NoStop}%
\bibitem [{\citenamefont {Fabrizio}(1996)}]{fabrizioSuperconductivityDopingSpinliquid1996}%
  \BibitemOpen
  \bibfield  {author} {\bibinfo {author} {\bibfnamefont {M.}~\bibnamefont {Fabrizio}},\ }\bibfield  {title} {\bibinfo {title} {Superconductivity from doping a spin-liquid insulator: {{A}} simple one-dimensional example},\ }\href {https://doi.org/10.1103/PhysRevB.54.10054} {\bibfield  {journal} {\bibinfo  {journal} {Phys. Rev. B}\ }\textbf {\bibinfo {volume} {54}},\ \bibinfo {pages} {10054} (\bibinfo {year} {1996})}\BibitemShut {NoStop}%
\bibitem [{\citenamefont {Konik}\ \emph {et~al.}(2006)\citenamefont {Konik}, \citenamefont {Rice},\ and\ \citenamefont {Tsvelik}}]{konikDopedSpinLiquid2006}%
  \BibitemOpen
  \bibfield  {author} {\bibinfo {author} {\bibfnamefont {R.~M.}\ \bibnamefont {Konik}}, \bibinfo {author} {\bibfnamefont {T.~M.}\ \bibnamefont {Rice}},\ and\ \bibinfo {author} {\bibfnamefont {A.~M.}\ \bibnamefont {Tsvelik}},\ }\bibfield  {title} {\bibinfo {title} {Doped {{Spin Liquid}}: {{Luttinger Sum Rule}} and {{Low Temperature Order}}},\ }\href {https://doi.org/10.1103/PhysRevLett.96.086407} {\bibfield  {journal} {\bibinfo  {journal} {Phys. Rev. Lett.}\ }\textbf {\bibinfo {volume} {96}},\ \bibinfo {pages} {086407} (\bibinfo {year} {2006})}\BibitemShut {NoStop}%
\bibitem [{\citenamefont {Rokhsar}(1993)}]{rokhsarPairingDopedSpin1993}%
  \BibitemOpen
  \bibfield  {author} {\bibinfo {author} {\bibfnamefont {D.~S.}\ \bibnamefont {Rokhsar}},\ }\bibfield  {title} {\bibinfo {title} {Pairing in doped spin liquids: {{Anyon}} versus d-wave superconductivity},\ }\href {https://doi.org/10.1103/PhysRevLett.70.493} {\bibfield  {journal} {\bibinfo  {journal} {Phys. Rev. Lett.}\ }\textbf {\bibinfo {volume} {70}},\ \bibinfo {pages} {493} (\bibinfo {year} {1993})}\BibitemShut {NoStop}%
\bibitem [{\citenamefont {Kelly}\ \emph {et~al.}(2016)\citenamefont {Kelly}, \citenamefont {Gallagher},\ and\ \citenamefont {McQueen}}]{kellyElectronDopingKagome2016}%
  \BibitemOpen
  \bibfield  {author} {\bibinfo {author} {\bibfnamefont {Z.~A.}\ \bibnamefont {Kelly}}, \bibinfo {author} {\bibfnamefont {M.~J.}\ \bibnamefont {Gallagher}},\ and\ \bibinfo {author} {\bibfnamefont {T.~M.}\ \bibnamefont {McQueen}},\ }\bibfield  {title} {\bibinfo {title} {Electron {{Doping}} a {{Kagome Spin Liquid}}},\ }\href {https://doi.org/10.1103/PhysRevX.6.041007} {\bibfield  {journal} {\bibinfo  {journal} {Phys. Rev. X}\ }\textbf {\bibinfo {volume} {6}},\ \bibinfo {pages} {041007} (\bibinfo {year} {2016})}\BibitemShut {NoStop}%
\bibitem [{\citenamefont {Senthil}\ and\ \citenamefont {Lee}(2005)}]{senthilCupratesDoped$U1$2005}%
  \BibitemOpen
  \bibfield  {author} {\bibinfo {author} {\bibfnamefont {T.}~\bibnamefont {Senthil}}\ and\ \bibinfo {author} {\bibfnamefont {P.~A.}\ \bibnamefont {Lee}},\ }\bibfield  {title} {\bibinfo {title} {Cuprates as doped \${{U}}(1)\$ spin liquids},\ }\href {https://doi.org/10.1103/PhysRevB.71.174515} {\bibfield  {journal} {\bibinfo  {journal} {Phys. Rev. B}\ }\textbf {\bibinfo {volume} {71}},\ \bibinfo {pages} {174515} (\bibinfo {year} {2005})}\BibitemShut {NoStop}%
\bibitem [{\citenamefont {Sigrist}\ \emph {et~al.}(1994)\citenamefont {Sigrist}, \citenamefont {Rice},\ and\ \citenamefont {Zhang}}]{sigristSuperconductivityQuasionedimensionalSpin1994}%
  \BibitemOpen
  \bibfield  {author} {\bibinfo {author} {\bibfnamefont {M.}~\bibnamefont {Sigrist}}, \bibinfo {author} {\bibfnamefont {T.~M.}\ \bibnamefont {Rice}},\ and\ \bibinfo {author} {\bibfnamefont {F.~C.}\ \bibnamefont {Zhang}},\ }\bibfield  {title} {\bibinfo {title} {Superconductivity in a quasi-one-dimensional spin liquid},\ }\href {https://doi.org/10.1103/PhysRevB.49.12058} {\bibfield  {journal} {\bibinfo  {journal} {Phys. Rev. B}\ }\textbf {\bibinfo {volume} {49}},\ \bibinfo {pages} {12058} (\bibinfo {year} {1994})}\BibitemShut {NoStop}%
\bibitem [{\citenamefont {Kalmeyer}\ and\ \citenamefont {Laughlin}(1987)}]{kalmeyer1987equivalence}%
  \BibitemOpen
  \bibfield  {author} {\bibinfo {author} {\bibfnamefont {V.}~\bibnamefont {Kalmeyer}}\ and\ \bibinfo {author} {\bibfnamefont {R.~B.}\ \bibnamefont {Laughlin}},\ }\bibfield  {title} {\bibinfo {title} {Equivalence of the resonating-valence-bond and fractional quantum hall states},\ }\href {https://doi.org/10.1103/PhysRevLett.59.2095} {\bibfield  {journal} {\bibinfo  {journal} {Phys. Rev. Lett.}\ }\textbf {\bibinfo {volume} {59}},\ \bibinfo {pages} {2095} (\bibinfo {year} {1987})}\BibitemShut {NoStop}%
\bibitem [{\citenamefont {Laughlin}(1988)}]{laughlin1988}%
  \BibitemOpen
  \bibfield  {author} {\bibinfo {author} {\bibfnamefont {R.~B.}\ \bibnamefont {Laughlin}},\ }\bibfield  {title} {\bibinfo {title} {Superconducting ground state of noninteracting particles obeying fractional statistics},\ }\href {https://doi.org/10.1103/PhysRevLett.60.2677} {\bibfield  {journal} {\bibinfo  {journal} {Phys. Rev. Lett.}\ }\textbf {\bibinfo {volume} {60}},\ \bibinfo {pages} {2677} (\bibinfo {year} {1988})}\BibitemShut {NoStop}%
\bibitem [{\citenamefont {Wen}\ \emph {et~al.}(1989)\citenamefont {Wen}, \citenamefont {Wilczek},\ and\ \citenamefont {Zee}}]{wen1989chiral}%
  \BibitemOpen
  \bibfield  {author} {\bibinfo {author} {\bibfnamefont {X.~G.}\ \bibnamefont {Wen}}, \bibinfo {author} {\bibfnamefont {F.}~\bibnamefont {Wilczek}},\ and\ \bibinfo {author} {\bibfnamefont {A.}~\bibnamefont {Zee}},\ }\bibfield  {title} {\bibinfo {title} {Chiral spin states and superconductivity},\ }\href {https://doi.org/10.1103/PhysRevB.39.11413} {\bibfield  {journal} {\bibinfo  {journal} {Phys. Rev. B}\ }\textbf {\bibinfo {volume} {39}},\ \bibinfo {pages} {11413} (\bibinfo {year} {1989})}\BibitemShut {NoStop}%
\bibitem [{\citenamefont {Lee}\ and\ \citenamefont {Fisher}(1989)}]{lee1989}%
  \BibitemOpen
  \bibfield  {author} {\bibinfo {author} {\bibfnamefont {D.-H.}\ \bibnamefont {Lee}}\ and\ \bibinfo {author} {\bibfnamefont {M.~P.~A.}\ \bibnamefont {Fisher}},\ }\bibfield  {title} {\bibinfo {title} {Anyon superconductivity and the fractional quantum hall effect},\ }\href {https://doi.org/10.1103/PhysRevLett.63.903} {\bibfield  {journal} {\bibinfo  {journal} {Phys. Rev. Lett.}\ }\textbf {\bibinfo {volume} {63}},\ \bibinfo {pages} {903} (\bibinfo {year} {1989})}\BibitemShut {NoStop}%
\bibitem [{\citenamefont {Balents}(2010)}]{balents2010spin}%
  \BibitemOpen
  \bibfield  {author} {\bibinfo {author} {\bibfnamefont {L.}~\bibnamefont {Balents}},\ }\bibfield  {title} {\bibinfo {title} {Spin liquids in frustrated magnets},\ }\href {https://www.nature.com/articles/nature08917} {\bibfield  {journal} {\bibinfo  {journal} {Nature}\ }\textbf {\bibinfo {volume} {464}},\ \bibinfo {pages} {199} (\bibinfo {year} {2010})}\BibitemShut {NoStop}%
\bibitem [{\citenamefont {He}\ \emph {et~al.}(2014)\citenamefont {He}, \citenamefont {Sheng},\ and\ \citenamefont {Chen}}]{he2014}%
  \BibitemOpen
  \bibfield  {author} {\bibinfo {author} {\bibfnamefont {Y.-C.}\ \bibnamefont {He}}, \bibinfo {author} {\bibfnamefont {D.~N.}\ \bibnamefont {Sheng}},\ and\ \bibinfo {author} {\bibfnamefont {Y.}~\bibnamefont {Chen}},\ }\bibfield  {title} {\bibinfo {title} {Chiral spin liquid in a frustrated anisotropic kagome heisenberg model},\ }\href {https://doi.org/10.1103/PhysRevLett.112.137202} {\bibfield  {journal} {\bibinfo  {journal} {Phys. Rev. Lett.}\ }\textbf {\bibinfo {volume} {112}},\ \bibinfo {pages} {137202} (\bibinfo {year} {2014})}\BibitemShut {NoStop}%
\bibitem [{\citenamefont {Gong}\ \emph {et~al.}(2014)\citenamefont {Gong}, \citenamefont {Zhu},\ and\ \citenamefont {Sheng}}]{gong2014}%
  \BibitemOpen
  \bibfield  {author} {\bibinfo {author} {\bibfnamefont {S.-S.}\ \bibnamefont {Gong}}, \bibinfo {author} {\bibfnamefont {W.}~\bibnamefont {Zhu}},\ and\ \bibinfo {author} {\bibfnamefont {D.~N.}\ \bibnamefont {Sheng}},\ }\bibfield  {title} {\bibinfo {title} {Emergent chiral spin liquid: Fractional quantum hall effect in a kagome heisenberg model},\ }\href {https://doi.org/10.1038/srep06317} {\bibfield  {journal} {\bibinfo  {journal} {Scientific Reports}\ }\textbf {\bibinfo {volume} {4}},\ \bibinfo {pages} {6317} (\bibinfo {year} {2014})}\BibitemShut {NoStop}%
\bibitem [{\citenamefont {Bauer}\ \emph {et~al.}(2014)\citenamefont {Bauer}, \citenamefont {Cincio}, \citenamefont {Keller}, \citenamefont {Dolfi}, \citenamefont {Vidal}, \citenamefont {Trebst},\ and\ \citenamefont {Ludwig}}]{bauer2014}%
  \BibitemOpen
  \bibfield  {author} {\bibinfo {author} {\bibfnamefont {B.}~\bibnamefont {Bauer}}, \bibinfo {author} {\bibfnamefont {L.}~\bibnamefont {Cincio}}, \bibinfo {author} {\bibfnamefont {B.~P.}\ \bibnamefont {Keller}}, \bibinfo {author} {\bibfnamefont {M.}~\bibnamefont {Dolfi}}, \bibinfo {author} {\bibfnamefont {G.}~\bibnamefont {Vidal}}, \bibinfo {author} {\bibfnamefont {S.}~\bibnamefont {Trebst}},\ and\ \bibinfo {author} {\bibfnamefont {A.~W.~W.}\ \bibnamefont {Ludwig}},\ }\bibfield  {title} {\bibinfo {title} {Chiral spin liquid and emergent anyons in a kagome lattice mott insulator},\ }\href {https://doi.org/10.1038/ncomms6137} {\bibfield  {journal} {\bibinfo  {journal} {Nature Communications}\ }\textbf {\bibinfo {volume} {5}},\ \bibinfo {pages} {5137} (\bibinfo {year} {2014})}\BibitemShut {NoStop}%
\bibitem [{\citenamefont {Gong}\ \emph {et~al.}(2015)\citenamefont {Gong}, \citenamefont {Zhu}, \citenamefont {Balents},\ and\ \citenamefont {Sheng}}]{gong2015}%
  \BibitemOpen
  \bibfield  {author} {\bibinfo {author} {\bibfnamefont {S.-S.}\ \bibnamefont {Gong}}, \bibinfo {author} {\bibfnamefont {W.}~\bibnamefont {Zhu}}, \bibinfo {author} {\bibfnamefont {L.}~\bibnamefont {Balents}},\ and\ \bibinfo {author} {\bibfnamefont {D.~N.}\ \bibnamefont {Sheng}},\ }\bibfield  {title} {\bibinfo {title} {Global phase diagram of competing ordered and quantum spin-liquid phases on the kagome lattice},\ }\href {https://doi.org/10.1103/PhysRevB.91.075112} {\bibfield  {journal} {\bibinfo  {journal} {Phys. Rev. B}\ }\textbf {\bibinfo {volume} {91}},\ \bibinfo {pages} {075112} (\bibinfo {year} {2015})}\BibitemShut {NoStop}%
\bibitem [{\citenamefont {Szasz}\ \emph {et~al.}(2020)\citenamefont {Szasz}, \citenamefont {Motruk}, \citenamefont {Zaletel},\ and\ \citenamefont {Moore}}]{szasz2020chiral}%
  \BibitemOpen
  \bibfield  {author} {\bibinfo {author} {\bibfnamefont {A.}~\bibnamefont {Szasz}}, \bibinfo {author} {\bibfnamefont {J.}~\bibnamefont {Motruk}}, \bibinfo {author} {\bibfnamefont {M.~P.}\ \bibnamefont {Zaletel}},\ and\ \bibinfo {author} {\bibfnamefont {J.~E.}\ \bibnamefont {Moore}},\ }\bibfield  {title} {\bibinfo {title} {Chiral spin liquid phase of the triangular lattice hubbard model: A density matrix renormalization group study},\ }\href {https://doi.org/10.1103/PhysRevX.10.021042} {\bibfield  {journal} {\bibinfo  {journal} {Phys. Rev. X}\ }\textbf {\bibinfo {volume} {10}},\ \bibinfo {pages} {021042} (\bibinfo {year} {2020})}\BibitemShut {NoStop}%
\bibitem [{\citenamefont {Chen}\ \emph {et~al.}(2022)\citenamefont {Chen}, \citenamefont {Chen}, \citenamefont {Gong}, \citenamefont {Sheng}, \citenamefont {Li},\ and\ \citenamefont {Weichselbaum}}]{chen2021quantum}%
  \BibitemOpen
  \bibfield  {author} {\bibinfo {author} {\bibfnamefont {B.-B.}\ \bibnamefont {Chen}}, \bibinfo {author} {\bibfnamefont {Z.}~\bibnamefont {Chen}}, \bibinfo {author} {\bibfnamefont {S.-S.}\ \bibnamefont {Gong}}, \bibinfo {author} {\bibfnamefont {D.~N.}\ \bibnamefont {Sheng}}, \bibinfo {author} {\bibfnamefont {W.}~\bibnamefont {Li}},\ and\ \bibinfo {author} {\bibfnamefont {A.}~\bibnamefont {Weichselbaum}},\ }\bibfield  {title} {\bibinfo {title} {Quantum spin liquid with emergent chiral order in the triangular-lattice hubbard model},\ }\href {https://doi.org/10.1103/PhysRevB.106.094420} {\bibfield  {journal} {\bibinfo  {journal} {Phys. Rev. B}\ }\textbf {\bibinfo {volume} {106}},\ \bibinfo {pages} {094420} (\bibinfo {year} {2022})}\BibitemShut {NoStop}%
\bibitem [{\citenamefont {Wietek}\ \emph {et~al.}(2021)\citenamefont {Wietek}, \citenamefont {Rossi}, \citenamefont {\ifmmode~\check{S}\else \v{S}\fi{}imkovic}, \citenamefont {Klett}, \citenamefont {Hansmann}, \citenamefont {Ferrero}, \citenamefont {Stoudenmire}, \citenamefont {Sch\"afer},\ and\ \citenamefont {Georges}}]{wietek2021}%
  \BibitemOpen
  \bibfield  {author} {\bibinfo {author} {\bibfnamefont {A.}~\bibnamefont {Wietek}}, \bibinfo {author} {\bibfnamefont {R.}~\bibnamefont {Rossi}}, \bibinfo {author} {\bibfnamefont {F.}~\bibnamefont {\ifmmode~\check{S}\else \v{S}\fi{}imkovic}}, \bibinfo {author} {\bibfnamefont {M.}~\bibnamefont {Klett}}, \bibinfo {author} {\bibfnamefont {P.}~\bibnamefont {Hansmann}}, \bibinfo {author} {\bibfnamefont {M.}~\bibnamefont {Ferrero}}, \bibinfo {author} {\bibfnamefont {E.~M.}\ \bibnamefont {Stoudenmire}}, \bibinfo {author} {\bibfnamefont {T.}~\bibnamefont {Sch\"afer}},\ and\ \bibinfo {author} {\bibfnamefont {A.}~\bibnamefont {Georges}},\ }\bibfield  {title} {\bibinfo {title} {Mott insulating states with competing orders in the triangular lattice hubbard model},\ }\href {https://doi.org/10.1103/PhysRevX.11.041013} {\bibfield  {journal} {\bibinfo  {journal} {Phys. Rev. X}\ }\textbf {\bibinfo {volume} {11}},\ \bibinfo {pages} {041013} (\bibinfo {year} {2021})}\BibitemShut {NoStop}%
\bibitem [{\citenamefont {Zhou}\ \emph {et~al.}(2022)\citenamefont {Zhou}, \citenamefont {Sheng},\ and\ \citenamefont {Kim}}]{zhou2022}%
  \BibitemOpen
  \bibfield  {author} {\bibinfo {author} {\bibfnamefont {Y.}~\bibnamefont {Zhou}}, \bibinfo {author} {\bibfnamefont {D.~N.}\ \bibnamefont {Sheng}},\ and\ \bibinfo {author} {\bibfnamefont {E.-A.}\ \bibnamefont {Kim}},\ }\bibfield  {title} {\bibinfo {title} {Quantum phases of transition metal dichalcogenide moir\'e systems},\ }\href {https://doi.org/10.1103/PhysRevLett.128.157602} {\bibfield  {journal} {\bibinfo  {journal} {Phys. Rev. Lett.}\ }\textbf {\bibinfo {volume} {128}},\ \bibinfo {pages} {157602} (\bibinfo {year} {2022})}\BibitemShut {NoStop}%
\bibitem [{\citenamefont {Peng}\ \emph {et~al.}(2021)\citenamefont {Peng}, \citenamefont {Jiang}, \citenamefont {Sheng},\ and\ \citenamefont {Jiang}}]{peng2021doping}%
  \BibitemOpen
  \bibfield  {author} {\bibinfo {author} {\bibfnamefont {C.}~\bibnamefont {Peng}}, \bibinfo {author} {\bibfnamefont {Y.-F.}\ \bibnamefont {Jiang}}, \bibinfo {author} {\bibfnamefont {D.-N.}\ \bibnamefont {Sheng}},\ and\ \bibinfo {author} {\bibfnamefont {H.-C.}\ \bibnamefont {Jiang}},\ }\bibfield  {title} {\bibinfo {title} {Doping quantum spin liquids on the kagome lattice},\ }\href {https://onlinelibrary.wiley.com/doi/abs/10.1002/qute.202000126} {\bibfield  {journal} {\bibinfo  {journal} {Advanced Quantum Technologies}\ }\textbf {\bibinfo {volume} {4}},\ \bibinfo {pages} {2000126} (\bibinfo {year} {2021})}\BibitemShut {NoStop}%
\bibitem [{\citenamefont {Zhu}\ \emph {et~al.}(2022)\citenamefont {Zhu}, \citenamefont {Sheng},\ and\ \citenamefont {Vishwanath}}]{zhu2022doped}%
  \BibitemOpen
  \bibfield  {author} {\bibinfo {author} {\bibfnamefont {Z.}~\bibnamefont {Zhu}}, \bibinfo {author} {\bibfnamefont {D.~N.}\ \bibnamefont {Sheng}},\ and\ \bibinfo {author} {\bibfnamefont {A.}~\bibnamefont {Vishwanath}},\ }\bibfield  {title} {\bibinfo {title} {Doped mott insulators in the triangular-lattice hubbard model},\ }\href {https://doi.org/10.1103/PhysRevB.105.205110} {\bibfield  {journal} {\bibinfo  {journal} {Phys. Rev. B}\ }\textbf {\bibinfo {volume} {105}},\ \bibinfo {pages} {205110} (\bibinfo {year} {2022})}\BibitemShut {NoStop}%
\bibitem [{\citenamefont {Jiang}\ and\ \citenamefont {Jiang}(2020)}]{jiang2020topological}%
  \BibitemOpen
  \bibfield  {author} {\bibinfo {author} {\bibfnamefont {Y.-F.}\ \bibnamefont {Jiang}}\ and\ \bibinfo {author} {\bibfnamefont {H.-C.}\ \bibnamefont {Jiang}},\ }\bibfield  {title} {\bibinfo {title} {Topological superconductivity in the doped chiral spin liquid on the triangular lattice},\ }\href {https://doi.org/10.1103/PhysRevLett.125.157002} {\bibfield  {journal} {\bibinfo  {journal} {Phys. Rev. Lett.}\ }\textbf {\bibinfo {volume} {125}},\ \bibinfo {pages} {157002} (\bibinfo {year} {2020})}\BibitemShut {NoStop}%
\bibitem [{\citenamefont {Huang}\ and\ \citenamefont {Sheng}(2022)}]{huang2021topological}%
  \BibitemOpen
  \bibfield  {author} {\bibinfo {author} {\bibfnamefont {Y.}~\bibnamefont {Huang}}\ and\ \bibinfo {author} {\bibfnamefont {D.~N.}\ \bibnamefont {Sheng}},\ }\bibfield  {title} {\bibinfo {title} {Topological chiral and nematic superconductivity by doping mott insulators on triangular lattice},\ }\href {https://doi.org/10.1103/PhysRevX.12.031009} {\bibfield  {journal} {\bibinfo  {journal} {Phys. Rev. X}\ }\textbf {\bibinfo {volume} {12}},\ \bibinfo {pages} {031009} (\bibinfo {year} {2022})}\BibitemShut {NoStop}%
\bibitem [{\citenamefont {Huang}\ \emph {et~al.}(2023)\citenamefont {Huang}, \citenamefont {Gong},\ and\ \citenamefont {Sheng}}]{huangQuantumPhaseDiagram2023c}%
  \BibitemOpen
  \bibfield  {author} {\bibinfo {author} {\bibfnamefont {Y.}~\bibnamefont {Huang}}, \bibinfo {author} {\bibfnamefont {S.-S.}\ \bibnamefont {Gong}},\ and\ \bibinfo {author} {\bibfnamefont {D.~N.}\ \bibnamefont {Sheng}},\ }\bibfield  {title} {\bibinfo {title} {Quantum {{Phase Diagram}} and {{Spontaneously Emergent Topological Chiral Superconductivity}} in {{Doped Triangular-Lattice Mott Insulators}}},\ }\href {https://doi.org/10.1103/PhysRevLett.130.136003} {\bibfield  {journal} {\bibinfo  {journal} {Phys. Rev. Lett.}\ }\textbf {\bibinfo {volume} {130}},\ \bibinfo {pages} {136003} (\bibinfo {year} {2023})}\BibitemShut {NoStop}%
\bibitem [{\citenamefont {Fradkin}\ \emph {et~al.}(2015)\citenamefont {Fradkin}, \citenamefont {Kivelson},\ and\ \citenamefont {Tranquada}}]{fradkin2015colloquium}%
  \BibitemOpen
  \bibfield  {author} {\bibinfo {author} {\bibfnamefont {E.}~\bibnamefont {Fradkin}}, \bibinfo {author} {\bibfnamefont {S.~A.}\ \bibnamefont {Kivelson}},\ and\ \bibinfo {author} {\bibfnamefont {J.~M.}\ \bibnamefont {Tranquada}},\ }\bibfield  {title} {\bibinfo {title} {Colloquium: Theory of intertwined orders in high temperature superconductors},\ }\href {https://doi.org/10.1103/RevModPhys.87.457} {\bibfield  {journal} {\bibinfo  {journal} {Rev. Mod. Phys.}\ }\textbf {\bibinfo {volume} {87}},\ \bibinfo {pages} {457} (\bibinfo {year} {2015})}\BibitemShut {NoStop}%
\bibitem [{\citenamefont {Song}\ \emph {et~al.}(2021)\citenamefont {Song}, \citenamefont {Vishwanath},\ and\ \citenamefont {Zhang}}]{song2021doping}%
  \BibitemOpen
  \bibfield  {author} {\bibinfo {author} {\bibfnamefont {X.-Y.}\ \bibnamefont {Song}}, \bibinfo {author} {\bibfnamefont {A.}~\bibnamefont {Vishwanath}},\ and\ \bibinfo {author} {\bibfnamefont {Y.-H.}\ \bibnamefont {Zhang}},\ }\bibfield  {title} {\bibinfo {title} {Doping the chiral spin liquid: Topological superconductor or chiral metal},\ }\href {https://doi.org/10.1103/PhysRevB.103.165138} {\bibfield  {journal} {\bibinfo  {journal} {Phys. Rev. B}\ }\textbf {\bibinfo {volume} {103}},\ \bibinfo {pages} {165138} (\bibinfo {year} {2021})}\BibitemShut {NoStop}%
\bibitem [{\citenamefont {Khatua}\ \emph {et~al.}(2023)\citenamefont {Khatua}, \citenamefont {Sana}, \citenamefont {Zorko}, \citenamefont {Gomil{\v s}ek}, \citenamefont {Sethupathi}, \citenamefont {Rao}, \citenamefont {Baenitz}, \citenamefont {Schmidt},\ and\ \citenamefont {Khuntia}}]{khatuaExperimentalSignaturesQuantum2023}%
  \BibitemOpen
  \bibfield  {author} {\bibinfo {author} {\bibfnamefont {J.}~\bibnamefont {Khatua}}, \bibinfo {author} {\bibfnamefont {B.}~\bibnamefont {Sana}}, \bibinfo {author} {\bibfnamefont {A.}~\bibnamefont {Zorko}}, \bibinfo {author} {\bibfnamefont {M.}~\bibnamefont {Gomil{\v s}ek}}, \bibinfo {author} {\bibfnamefont {K.}~\bibnamefont {Sethupathi}}, \bibinfo {author} {\bibfnamefont {M.~S.~R.}\ \bibnamefont {Rao}}, \bibinfo {author} {\bibfnamefont {M.}~\bibnamefont {Baenitz}}, \bibinfo {author} {\bibfnamefont {B.}~\bibnamefont {Schmidt}},\ and\ \bibinfo {author} {\bibfnamefont {P.}~\bibnamefont {Khuntia}},\ }\bibfield  {title} {\bibinfo {title} {Experimental signatures of quantum and topological states in frustrated magnetism},\ }\href {https://doi.org/10.1016/j.physrep.2023.09.008} {\bibfield  {journal} {\bibinfo  {journal} {Physics Reports}\ }\bibinfo {series} {Experimental Signatures of Quantum and Topological States in Frustrated Magnetism},\ \textbf {\bibinfo {volume} {1041}},\ \bibinfo {pages} {1} (\bibinfo {year}
  {2023})}\BibitemShut {NoStop}%
\bibitem [{\citenamefont {Knolle}\ and\ \citenamefont {Moessner}(2019)}]{knolleFieldGuideSpin2019}%
  \BibitemOpen
  \bibfield  {author} {\bibinfo {author} {\bibfnamefont {J.}~\bibnamefont {Knolle}}\ and\ \bibinfo {author} {\bibfnamefont {R.}~\bibnamefont {Moessner}},\ }\bibfield  {title} {\bibinfo {title} {A {{Field Guide}} to {{Spin Liquids}}},\ }\href {https://doi.org/10.1146/annurev-conmatphys-031218-013401} {\bibfield  {journal} {\bibinfo  {journal} {Annual Review of Condensed Matter Physics}\ }\textbf {\bibinfo {volume} {10}},\ \bibinfo {pages} {451} (\bibinfo {year} {2019})}\BibitemShut {NoStop}%
\bibitem [{\citenamefont {Wu}\ \emph {et~al.}(2018)\citenamefont {Wu}, \citenamefont {Lovorn}, \citenamefont {Tutuc},\ and\ \citenamefont {MacDonald}}]{wu2018hubbard}%
  \BibitemOpen
  \bibfield  {author} {\bibinfo {author} {\bibfnamefont {F.}~\bibnamefont {Wu}}, \bibinfo {author} {\bibfnamefont {T.}~\bibnamefont {Lovorn}}, \bibinfo {author} {\bibfnamefont {E.}~\bibnamefont {Tutuc}},\ and\ \bibinfo {author} {\bibfnamefont {A.~H.}\ \bibnamefont {MacDonald}},\ }\bibfield  {title} {\bibinfo {title} {Hubbard model physics in transition metal dichalcogenide moir{\'e} bands},\ }\href {https://journals.aps.org/prl/abstract/10.1103/PhysRevLett.121.026402} {\bibfield  {journal} {\bibinfo  {journal} {Physical review letters}\ }\textbf {\bibinfo {volume} {121}},\ \bibinfo {pages} {026402} (\bibinfo {year} {2018})}\BibitemShut {NoStop}%
\bibitem [{\citenamefont {Tang}\ \emph {et~al.}(2020)\citenamefont {Tang}, \citenamefont {Li}, \citenamefont {Li}, \citenamefont {Xu}, \citenamefont {Liu}, \citenamefont {Barmak}, \citenamefont {Watanabe}, \citenamefont {Taniguchi}, \citenamefont {MacDonald}, \citenamefont {Shan},\ and\ \citenamefont {Mak}}]{Tang2020}%
  \BibitemOpen
  \bibfield  {author} {\bibinfo {author} {\bibfnamefont {Y.}~\bibnamefont {Tang}}, \bibinfo {author} {\bibfnamefont {L.}~\bibnamefont {Li}}, \bibinfo {author} {\bibfnamefont {T.}~\bibnamefont {Li}}, \bibinfo {author} {\bibfnamefont {Y.}~\bibnamefont {Xu}}, \bibinfo {author} {\bibfnamefont {S.}~\bibnamefont {Liu}}, \bibinfo {author} {\bibfnamefont {K.}~\bibnamefont {Barmak}}, \bibinfo {author} {\bibfnamefont {K.}~\bibnamefont {Watanabe}}, \bibinfo {author} {\bibfnamefont {T.}~\bibnamefont {Taniguchi}}, \bibinfo {author} {\bibfnamefont {A.~H.}\ \bibnamefont {MacDonald}}, \bibinfo {author} {\bibfnamefont {J.}~\bibnamefont {Shan}},\ and\ \bibinfo {author} {\bibfnamefont {K.~F.}\ \bibnamefont {Mak}},\ }\bibfield  {title} {\bibinfo {title} {Simulation of hubbard model physics in wse2/ws2 moir{\'e} superlattices},\ }\href {https://doi.org/10.1038/s41586-020-2085-3} {\bibfield  {journal} {\bibinfo  {journal} {Nature}\ }\textbf {\bibinfo {volume} {579}},\ \bibinfo {pages} {353} (\bibinfo {year} {2020})}\BibitemShut
  {NoStop}%
\bibitem [{\citenamefont {Kuhlenkamp}\ \emph {et~al.}(2024)\citenamefont {Kuhlenkamp}, \citenamefont {Kadow}, \citenamefont {Imamo{\u g}lu},\ and\ \citenamefont {Knap}}]{kuhlenkampChiralPseudospinLiquids2024b}%
  \BibitemOpen
  \bibfield  {author} {\bibinfo {author} {\bibfnamefont {C.}~\bibnamefont {Kuhlenkamp}}, \bibinfo {author} {\bibfnamefont {W.}~\bibnamefont {Kadow}}, \bibinfo {author} {\bibfnamefont {A.}~\bibnamefont {Imamo{\u g}lu}},\ and\ \bibinfo {author} {\bibfnamefont {M.}~\bibnamefont {Knap}},\ }\bibfield  {title} {\bibinfo {title} {Chiral {{Pseudospin Liquids}} in {{Moir}}{\textbackslash}'e {{Heterostructures}}},\ }\href {https://doi.org/10.1103/PhysRevX.14.021013} {\bibfield  {journal} {\bibinfo  {journal} {Phys. Rev. X}\ }\textbf {\bibinfo {volume} {14}},\ \bibinfo {pages} {021013} (\bibinfo {year} {2024})}\BibitemShut {NoStop}%
\bibitem [{\citenamefont {Divic}\ \emph {et~al.}(2024{\natexlab{a}})\citenamefont {Divic}, \citenamefont {Soejima}, \citenamefont {Crépel}, \citenamefont {Zaletel},\ and\ \citenamefont {Millis}}]{divic2024chiralspinliquidquantum}%
  \BibitemOpen
  \bibfield  {author} {\bibinfo {author} {\bibfnamefont {S.}~\bibnamefont {Divic}}, \bibinfo {author} {\bibfnamefont {T.}~\bibnamefont {Soejima}}, \bibinfo {author} {\bibfnamefont {V.}~\bibnamefont {Crépel}}, \bibinfo {author} {\bibfnamefont {M.~P.}\ \bibnamefont {Zaletel}},\ and\ \bibinfo {author} {\bibfnamefont {A.}~\bibnamefont {Millis}},\ }\href {https://arxiv.org/abs/2406.15348} {\bibinfo {title} {Chiral spin liquid and quantum phase transition in the triangular lattice hofstadter-hubbard model}} (\bibinfo {year} {2024}{\natexlab{a}}),\ \Eprint {https://arxiv.org/abs/2406.15348} {arXiv:2406.15348 [cond-mat.str-el]} \BibitemShut {NoStop}%
\bibitem [{\citenamefont {Divic}\ \emph {et~al.}(2024{\natexlab{b}})\citenamefont {Divic}, \citenamefont {Crépel}, \citenamefont {Soejima}, \citenamefont {Song}, \citenamefont {Millis}, \citenamefont {Zaletel},\ and\ \citenamefont {Vishwanath}}]{divic2024anyonsuperconductivitytopologicalcriticality}%
  \BibitemOpen
  \bibfield  {author} {\bibinfo {author} {\bibfnamefont {S.}~\bibnamefont {Divic}}, \bibinfo {author} {\bibfnamefont {V.}~\bibnamefont {Crépel}}, \bibinfo {author} {\bibfnamefont {T.}~\bibnamefont {Soejima}}, \bibinfo {author} {\bibfnamefont {X.-Y.}\ \bibnamefont {Song}}, \bibinfo {author} {\bibfnamefont {A.}~\bibnamefont {Millis}}, \bibinfo {author} {\bibfnamefont {M.~P.}\ \bibnamefont {Zaletel}},\ and\ \bibinfo {author} {\bibfnamefont {A.}~\bibnamefont {Vishwanath}},\ }\href {https://arxiv.org/abs/2410.18175} {\bibinfo {title} {Anyon superconductivity from topological criticality in a hofstadter-hubbard model}} (\bibinfo {year} {2024}{\natexlab{b}}),\ \Eprint {https://arxiv.org/abs/2410.18175} {arXiv:2410.18175 [cond-mat.str-el]} \BibitemShut {NoStop}%
\bibitem [{\citenamefont {Pichler}\ \emph {et~al.}(2025)\citenamefont {Pichler}, \citenamefont {Kuhlenkamp}, \citenamefont {Knap},\ and\ \citenamefont {Vishwanath}}]{pichler2025microscopicmechanismanyonsuperconductivity}%
  \BibitemOpen
  \bibfield  {author} {\bibinfo {author} {\bibfnamefont {F.}~\bibnamefont {Pichler}}, \bibinfo {author} {\bibfnamefont {C.}~\bibnamefont {Kuhlenkamp}}, \bibinfo {author} {\bibfnamefont {M.}~\bibnamefont {Knap}},\ and\ \bibinfo {author} {\bibfnamefont {A.}~\bibnamefont {Vishwanath}},\ }\href {https://arxiv.org/abs/2506.08000} {\bibinfo {title} {Microscopic mechanism of anyon superconductivity emerging from fractional chern insulators}} (\bibinfo {year} {2025}),\ \Eprint {https://arxiv.org/abs/2506.08000} {arXiv:2506.08000 [cond-mat.str-el]} \BibitemShut {NoStop}%
\bibitem [{\citenamefont {White}(1992)}]{white1992density}%
  \BibitemOpen
  \bibfield  {author} {\bibinfo {author} {\bibfnamefont {S.~R.}\ \bibnamefont {White}},\ }\bibfield  {title} {\bibinfo {title} {Density matrix formulation for quantum renormalization groups},\ }\href {https://journals.aps.org/prl/abstract/10.1103/PhysRevLett.69.2863} {\bibfield  {journal} {\bibinfo  {journal} {Physical review letters}\ }\textbf {\bibinfo {volume} {69}},\ \bibinfo {pages} {2863} (\bibinfo {year} {1992})}\BibitemShut {NoStop}%
\bibitem [{\citenamefont {Blankenbecler}\ \emph {et~al.}(1981)\citenamefont {Blankenbecler}, \citenamefont {Scalapino},\ and\ \citenamefont {Sugar}}]{DQMC1}%
  \BibitemOpen
  \bibfield  {author} {\bibinfo {author} {\bibfnamefont {R.}~\bibnamefont {Blankenbecler}}, \bibinfo {author} {\bibfnamefont {D.~J.}\ \bibnamefont {Scalapino}},\ and\ \bibinfo {author} {\bibfnamefont {R.~L.}\ \bibnamefont {Sugar}},\ }\bibfield  {title} {\bibinfo {title} {\uppercase{M}onte \uppercase{C}arlo calculations of coupled boson-fermion systems. {I}},\ }\href {https://doi.org/10.1103/PhysRevD.24.2278} {\bibfield  {journal} {\bibinfo  {journal} {Phys. Rev. D}\ }\textbf {\bibinfo {volume} {24}},\ \bibinfo {pages} {2278} (\bibinfo {year} {1981})}\BibitemShut {NoStop}%
\bibitem [{\citenamefont {White}\ \emph {et~al.}(1989)\citenamefont {White}, \citenamefont {Scalapino}, \citenamefont {Sugar}, \citenamefont {Loh}, \citenamefont {Gubernatis},\ and\ \citenamefont {Scalettar}}]{DQMC2}%
  \BibitemOpen
  \bibfield  {author} {\bibinfo {author} {\bibfnamefont {S.~R.}\ \bibnamefont {White}}, \bibinfo {author} {\bibfnamefont {D.~J.}\ \bibnamefont {Scalapino}}, \bibinfo {author} {\bibfnamefont {R.~L.}\ \bibnamefont {Sugar}}, \bibinfo {author} {\bibfnamefont {E.~Y.}\ \bibnamefont {Loh}}, \bibinfo {author} {\bibfnamefont {J.~E.}\ \bibnamefont {Gubernatis}},\ and\ \bibinfo {author} {\bibfnamefont {R.~T.}\ \bibnamefont {Scalettar}},\ }\bibfield  {title} {\bibinfo {title} {Numerical study of the two-dimensional \uppercase{H}ubbard model},\ }\href {https://doi.org/10.1103/PhysRevB.40.506} {\bibfield  {journal} {\bibinfo  {journal} {Phys. Rev. B}\ }\textbf {\bibinfo {volume} {40}},\ \bibinfo {pages} {506} (\bibinfo {year} {1989})}\BibitemShut {NoStop}%
\bibitem [{Sup()}]{SuppMaterial}%
  \BibitemOpen
  \href@noop {} {}\bibinfo {howpublished} {See Supplemental Materials at [URL will be inserted by publisher] for detailed numerical results and discussions.}\BibitemShut {Stop}%
\bibitem [{\citenamefont {McCulloch}(2007)}]{McCulloch2007}%
  \BibitemOpen
  \bibfield  {author} {\bibinfo {author} {\bibfnamefont {I.~P.}\ \bibnamefont {McCulloch}},\ }\bibfield  {title} {\bibinfo {title} {From density-matrix renormalization group to matrix product states},\ }\href {https://doi.org/10.1088/1742-5468/2007/10/p10014} {\bibfield  {journal} {\bibinfo  {journal} {Journal of Statistical Mechanics: Theory and Experiment}\ }\textbf {\bibinfo {volume} {2007}},\ \bibinfo {pages} {P10014} (\bibinfo {year} {2007})}\BibitemShut {NoStop}%
\bibitem [{\citenamefont {Shaffer}\ \emph {et~al.}(2022)\citenamefont {Shaffer}, \citenamefont {Wang},\ and\ \citenamefont {Santos}}]{shafferUnconventionalSelfsimilarHofstadter2022}%
  \BibitemOpen
  \bibfield  {author} {\bibinfo {author} {\bibfnamefont {D.}~\bibnamefont {Shaffer}}, \bibinfo {author} {\bibfnamefont {J.}~\bibnamefont {Wang}},\ and\ \bibinfo {author} {\bibfnamefont {L.~H.}\ \bibnamefont {Santos}},\ }\bibfield  {title} {\bibinfo {title} {Unconventional self-similar {{Hofstadter}} superconductivity from repulsive interactions},\ }\href {https://doi.org/10.1038/s41467-022-35316-z} {\bibfield  {journal} {\bibinfo  {journal} {Nat Commun}\ }\textbf {\bibinfo {volume} {13}},\ \bibinfo {pages} {7785} (\bibinfo {year} {2022})}\BibitemShut {NoStop}%
\bibitem [{\citenamefont {Shaffer}\ \emph {et~al.}(2021)\citenamefont {Shaffer}, \citenamefont {Wang},\ and\ \citenamefont {Santos}}]{shafferTheoryHofstadterSuperconductors2021}%
  \BibitemOpen
  \bibfield  {author} {\bibinfo {author} {\bibfnamefont {D.}~\bibnamefont {Shaffer}}, \bibinfo {author} {\bibfnamefont {J.}~\bibnamefont {Wang}},\ and\ \bibinfo {author} {\bibfnamefont {L.~H.}\ \bibnamefont {Santos}},\ }\bibfield  {title} {\bibinfo {title} {Theory of {{Hofstadter}} superconductors},\ }\href {https://doi.org/10.1103/PhysRevB.104.184501} {\bibfield  {journal} {\bibinfo  {journal} {Phys. Rev. B}\ }\textbf {\bibinfo {volume} {104}},\ \bibinfo {pages} {184501} (\bibinfo {year} {2021})}\BibitemShut {NoStop}%
\bibitem [{\citenamefont {Schollw{\"o}ck}(2011)}]{schollwock2011density}%
  \BibitemOpen
  \bibfield  {author} {\bibinfo {author} {\bibfnamefont {U.}~\bibnamefont {Schollw{\"o}ck}},\ }\bibfield  {title} {\bibinfo {title} {The density-matrix renormalization group in the age of matrix product states},\ }\href@noop {} {\bibfield  {journal} {\bibinfo  {journal} {Annals of physics}\ }\textbf {\bibinfo {volume} {326}},\ \bibinfo {pages} {96} (\bibinfo {year} {2011})}\BibitemShut {NoStop}%
\bibitem [{\citenamefont {Jiang}(2021)}]{jiang2021superconductivity}%
  \BibitemOpen
  \bibfield  {author} {\bibinfo {author} {\bibfnamefont {H.-C.}\ \bibnamefont {Jiang}},\ }\bibfield  {title} {\bibinfo {title} {Superconductivity in the doped quantum spin liquid on the triangular lattice},\ }\href {https://www.nature.com/articles/s41535-021-00375-w} {\bibfield  {journal} {\bibinfo  {journal} {npj Quantum Materials}\ }\textbf {\bibinfo {volume} {6}},\ \bibinfo {pages} {1} (\bibinfo {year} {2021})}\BibitemShut {NoStop}%
\bibitem [{\citenamefont {Zaletel}\ \emph {et~al.}(2014)\citenamefont {Zaletel}, \citenamefont {Mong},\ and\ \citenamefont {Pollmann}}]{zaletel2014flux}%
  \BibitemOpen
  \bibfield  {author} {\bibinfo {author} {\bibfnamefont {M.~P.}\ \bibnamefont {Zaletel}}, \bibinfo {author} {\bibfnamefont {R.~S.}\ \bibnamefont {Mong}},\ and\ \bibinfo {author} {\bibfnamefont {F.}~\bibnamefont {Pollmann}},\ }\bibfield  {title} {\bibinfo {title} {Flux insertion, entanglement, and quantized responses},\ }\href@noop {} {\bibfield  {journal} {\bibinfo  {journal} {Journal of Statistical Mechanics: Theory and Experiment}\ }\textbf {\bibinfo {volume} {2014}},\ \bibinfo {pages} {P10007} (\bibinfo {year} {2014})}\BibitemShut {NoStop}%
\bibitem [{\citenamefont {Cr\'epel}\ \emph {et~al.}(2022)\citenamefont {Cr\'epel}, \citenamefont {Cea}, \citenamefont {Fu},\ and\ \citenamefont {Guinea}}]{PhysRevB.105.094506}%
  \BibitemOpen
  \bibfield  {author} {\bibinfo {author} {\bibfnamefont {V.}~\bibnamefont {Cr\'epel}}, \bibinfo {author} {\bibfnamefont {T.}~\bibnamefont {Cea}}, \bibinfo {author} {\bibfnamefont {L.}~\bibnamefont {Fu}},\ and\ \bibinfo {author} {\bibfnamefont {F.}~\bibnamefont {Guinea}},\ }\bibfield  {title} {\bibinfo {title} {Unconventional superconductivity due to interband polarization},\ }\href {https://doi.org/10.1103/PhysRevB.105.094506} {\bibfield  {journal} {\bibinfo  {journal} {Phys. Rev. B}\ }\textbf {\bibinfo {volume} {105}},\ \bibinfo {pages} {094506} (\bibinfo {year} {2022})}\BibitemShut {NoStop}%
\bibitem [{\citenamefont {Markov}\ \emph {et~al.}(2019)\citenamefont {Markov}, \citenamefont {Rohringer},\ and\ \citenamefont {Rubtsov}}]{PhysRevB.100.115102}%
  \BibitemOpen
  \bibfield  {author} {\bibinfo {author} {\bibfnamefont {A.~A.}\ \bibnamefont {Markov}}, \bibinfo {author} {\bibfnamefont {G.}~\bibnamefont {Rohringer}},\ and\ \bibinfo {author} {\bibfnamefont {A.~N.}\ \bibnamefont {Rubtsov}},\ }\bibfield  {title} {\bibinfo {title} {Robustness of the topological quantization of the hall conductivity for correlated lattice electrons at finite temperatures},\ }\href {https://doi.org/10.1103/PhysRevB.100.115102} {\bibfield  {journal} {\bibinfo  {journal} {Phys. Rev. B}\ }\textbf {\bibinfo {volume} {100}},\ \bibinfo {pages} {115102} (\bibinfo {year} {2019})}\BibitemShut {NoStop}%
\bibitem [{\citenamefont {Kou}\ \emph {et~al.}(2015)\citenamefont {Kou}, \citenamefont {Pan}, \citenamefont {Wang}, \citenamefont {Fan}, \citenamefont {Choi}, \citenamefont {Lee}, \citenamefont {Nie}, \citenamefont {Murata}, \citenamefont {Shao}, \citenamefont {Zhang},\ and\ \citenamefont {Wang}}]{MIKou}%
  \BibitemOpen
  \bibfield  {author} {\bibinfo {author} {\bibfnamefont {X.}~\bibnamefont {Kou}}, \bibinfo {author} {\bibfnamefont {L.}~\bibnamefont {Pan}}, \bibinfo {author} {\bibfnamefont {J.}~\bibnamefont {Wang}}, \bibinfo {author} {\bibfnamefont {Y.}~\bibnamefont {Fan}}, \bibinfo {author} {\bibfnamefont {E.}~\bibnamefont {Choi}}, \bibinfo {author} {\bibfnamefont {W.-L.}\ \bibnamefont {Lee}}, \bibinfo {author} {\bibfnamefont {T.}~\bibnamefont {Nie}}, \bibinfo {author} {\bibfnamefont {K.}~\bibnamefont {Murata}}, \bibinfo {author} {\bibfnamefont {Q.}~\bibnamefont {Shao}}, \bibinfo {author} {\bibfnamefont {S.-C.}\ \bibnamefont {Zhang}},\ and\ \bibinfo {author} {\bibfnamefont {K.}~\bibnamefont {Wang}},\ }\bibfield  {title} {\bibinfo {title} {Metal-to-insulator switching in quantum anomalous hall states},\ }\href {https://doi.org/10.1038/ncomms9474} {\bibfield  {journal} {\bibinfo  {journal} {Nature Communications}\ }\textbf {\bibinfo {volume} {6}},\ \bibinfo {pages} {8474} (\bibinfo {year} {2015})}\BibitemShut {NoStop}%
\bibitem [{\citenamefont {Shi}\ and\ \citenamefont {Senthil}(2024)}]{shi2024dopingfractionalquantumanomalous}%
  \BibitemOpen
  \bibfield  {author} {\bibinfo {author} {\bibfnamefont {Z.~D.}\ \bibnamefont {Shi}}\ and\ \bibinfo {author} {\bibfnamefont {T.}~\bibnamefont {Senthil}},\ }\href {https://arxiv.org/abs/2409.20567} {\bibinfo {title} {Doping a fractional quantum anomalous hall insulator}} (\bibinfo {year} {2024}),\ \Eprint {https://arxiv.org/abs/2409.20567} {arXiv:2409.20567 [cond-mat.str-el]} \BibitemShut {NoStop}%
\bibitem [{\citenamefont {Kim}\ \emph {et~al.}(2025)\citenamefont {Kim}, \citenamefont {Timmel}, \citenamefont {Ju},\ and\ \citenamefont {Wen}}]{PhysRevB.111.014508}%
  \BibitemOpen
  \bibfield  {author} {\bibinfo {author} {\bibfnamefont {M.}~\bibnamefont {Kim}}, \bibinfo {author} {\bibfnamefont {A.}~\bibnamefont {Timmel}}, \bibinfo {author} {\bibfnamefont {L.}~\bibnamefont {Ju}},\ and\ \bibinfo {author} {\bibfnamefont {X.-G.}\ \bibnamefont {Wen}},\ }\bibfield  {title} {\bibinfo {title} {Topological chiral superconductivity beyond pairing in a fermi liquid},\ }\href {https://doi.org/10.1103/PhysRevB.111.014508} {\bibfield  {journal} {\bibinfo  {journal} {Phys. Rev. B}\ }\textbf {\bibinfo {volume} {111}},\ \bibinfo {pages} {014508} (\bibinfo {year} {2025})}\BibitemShut {NoStop}%
\bibitem [{\citenamefont {Shi}\ and\ \citenamefont {Senthil}(2025)}]{shi2025anyondelocalizationtransitionsdisordered}%
  \BibitemOpen
  \bibfield  {author} {\bibinfo {author} {\bibfnamefont {Z.~D.}\ \bibnamefont {Shi}}\ and\ \bibinfo {author} {\bibfnamefont {T.}~\bibnamefont {Senthil}},\ }\href {https://arxiv.org/abs/2506.02128} {\bibinfo {title} {Anyon delocalization transitions out of a disordered fqah insulator}} (\bibinfo {year} {2025}),\ \Eprint {https://arxiv.org/abs/2506.02128} {arXiv:2506.02128 [cond-mat.str-el]} \BibitemShut {NoStop}%
\bibitem [{pre()}]{preprint}%
  \BibitemOpen
  \href@noop {} {}\bibinfo {howpublished} {Clemens Kuhlenkamp, Stefan Divic, Michael P. Zaletel, Tomohiro Soejima, and Ashvin Vishwanath, Robust superconductivity upon doping chiral spin liquid and Chern insulators in a Hubbard-Hofstadter model. To appear.}\BibitemShut {Stop}%
\end{thebibliography}%

\clearpage
\appendix
\widetext
\begin{center}
\textbf{\large Supplemental Materials for: ``Chiral Topological Superconductivity in the Triangular-Lattice Hofstadter-Hubbard Model''}
\end{center}
\vspace{1mm}
\renewcommand\thefigure{\thesection S\arabic{figure}}
\renewcommand\theequation{\thesection S\arabic{equation}}
\setcounter{figure}{0} 
\setcounter{equation}{0} 

In the Supplemental Materials, we provide additional results to support the conclusions we have presented in the main text. In Sec. \label {sec: Symmetry}A, 
 we discuss pairing symmetry and its relation to magnetic translational symmetry, and present results of  symmetry of longer range pairing.  
In Sec. \label {sec:sc_cor}B, we provide the DMRG results of singlet pairing, spin, and density correlations and  the electronic Green's function
for more parameter points  on  ladders with $L_y=3, 4, 6, 8$ at hole doping levels $\delta =1/24,  1/12$ and $1/8$ to provide more
complete pictures of the emergent superconducting phases. We also present results of the spin chiral order at different $U$s.
In Sec. C, additional results on the electron and spin structure factors are provided for various Hubbard interactions.
In Sec. D and E, we present numerical results of the DQMC and a justification of using the correlation function at a nonzero imaginary time $\tau=
\beta/2$ in DQMC.  
In Sec. F, we present results based on the perturbation theory.

\begin{figure}
\centering
\includegraphics[width=0.8\textwidth,angle=0]{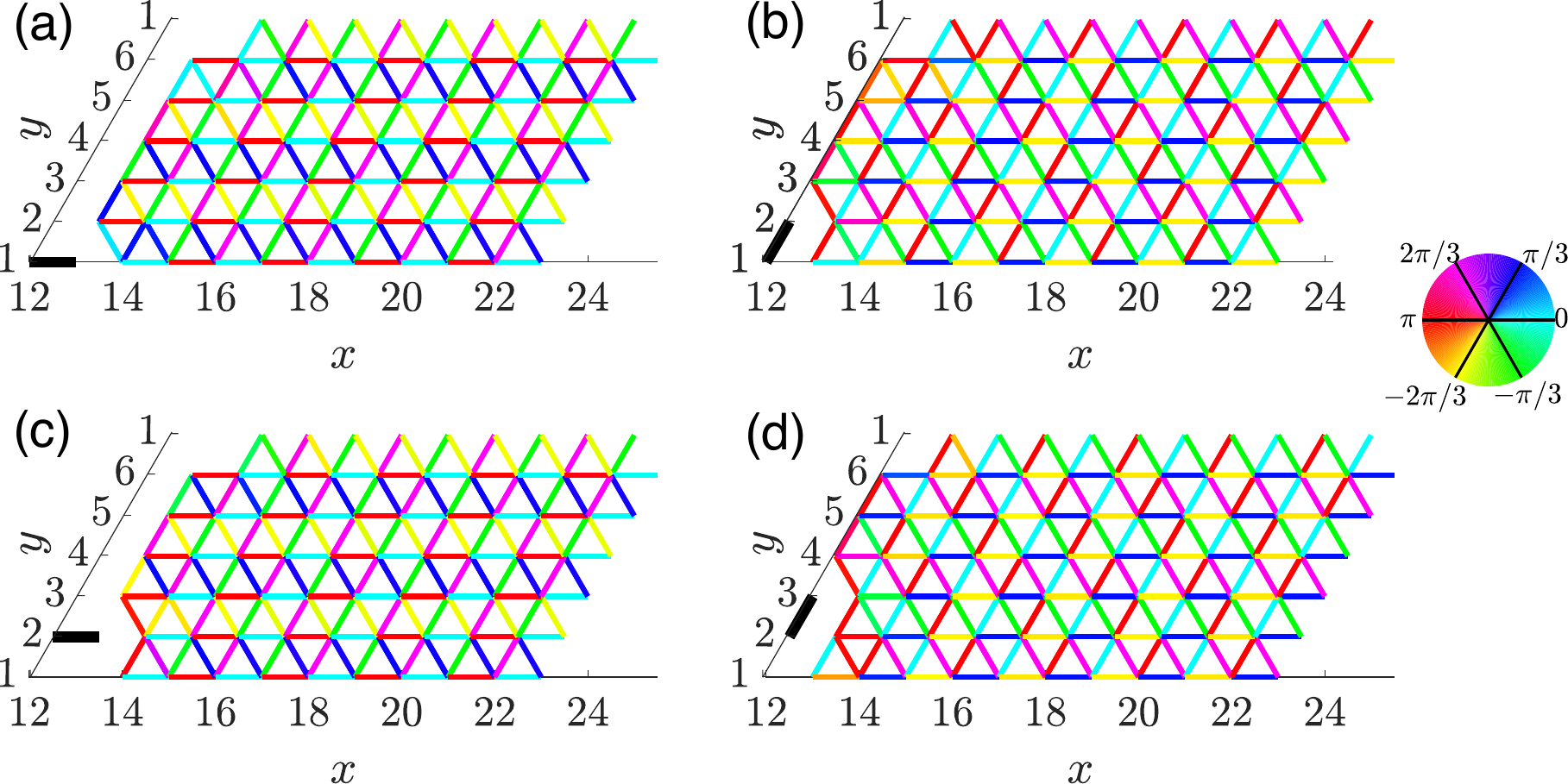}
   \caption{\label{symm_bond}Pairing phase structures when different reference bonds are chosen. Black bond lines represent the reference bonds.}
\end{figure}

\begin{figure}
\includegraphics[width=0.6\textwidth,angle=0]{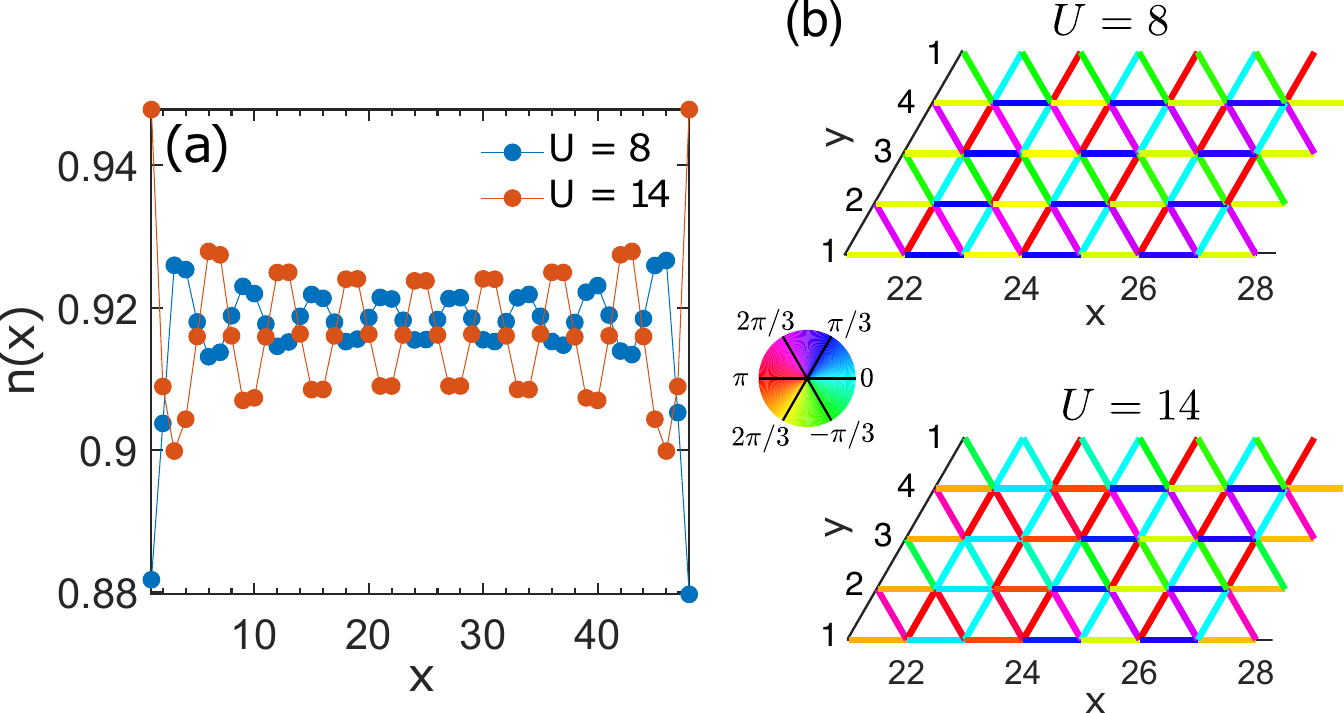}
   \caption{\label{Ne_symm_Y4} (a) Electron density profiles. (b) Phase structure of the pairing orders for $N=48\times 4, \delta=1/12$ and $M=12000$ for $U=8$ and 14.}
\end{figure}

\begin{figure}
\includegraphics[width=0.95\textwidth,angle=0]{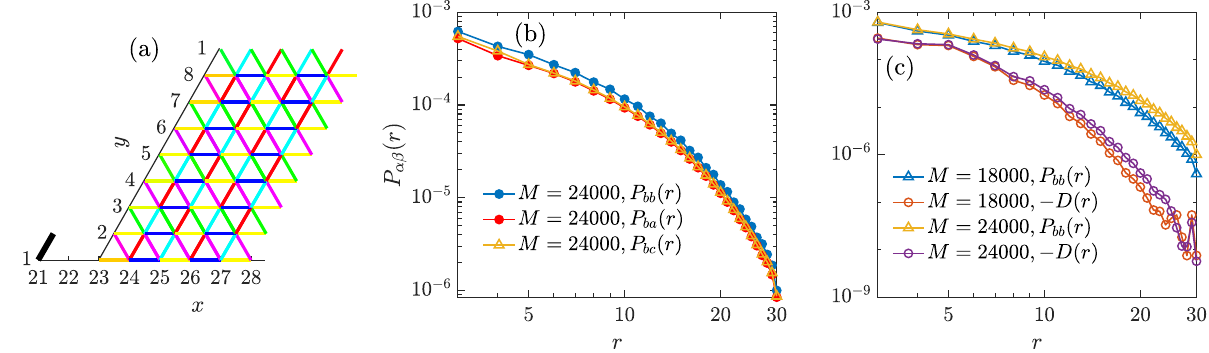}
   \caption{\label{Y8} (a) Pairing phase structure. The black bond at $x=21$ is the reference. (b) Pair-pair correlations $P_{\alpha\beta}(r)$ for different bonds. (c) Pair-pair and density-density $D(r)$ correlations. Parameters: $N=48\times 8, U=14, \delta=1/8$ and $M=18000$ and 24000.}
\end{figure}

\begin{figure}
\centering
\includegraphics[width=0.9\textwidth,angle=0]{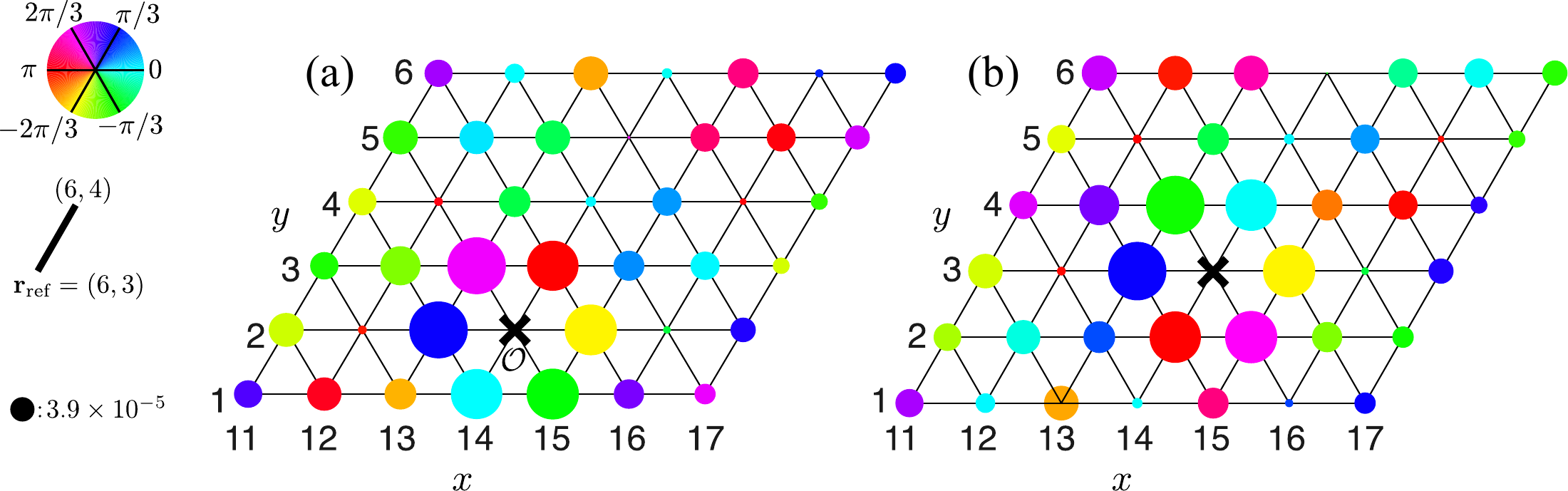}
   \caption{\label{Sc_long} The amplitudes and phases of the longer-range pair-pair correlation $\langle \hat{\Delta}^\dagger_a(\mathbf{r}_\text{ref}) \hat{\Delta}(\mathbf{r_0,r})\rangle$, where $\hat{\Delta}(\mathbf{r_0,r})=\langle\hat{c}_{\mathbf{r_0}\uparrow}\hat{c}_{\mathbf{r}\downarrow}-\hat{c}_{\mathbf{r_0}\downarrow}\hat{c}_{\mathbf{r}\uparrow} \rangle/\sqrt{2}$, $\mathbf{r_0}=(14,2)$ or $(14,3)$ is denoted by a cross in the figure and $\mathbf{r}$ represents the position of the pair partner. The area of circle is proportional to the amplitude of the correlation. The reference bond is shown on the left with $\mathbf{r}_\text{ref}=(6,3)$.}
\end{figure}


\section{\label{sec: Symmetry}A. Pairing Symmetry and  Magnetic Translation in the Imaginary $C_{6z}$ Gauge}

By inspecting the hopping phase structure in Fig.~\ref{Latt}, one can deduce the actions of the magnetic translation $\hat{T}_{a,b}$ along $\mathbf{e_{a,b}}$: 
\begin{equation}
    \hat{T}_a \hat{c}_{(x,y),\sigma} \hat{T}_a^{-1}=i(-1)^{x+y+1}\hat{c}_{(x+1,y),\sigma},\qquad \hat{T}_b \hat{c}_{(x,y),\sigma} \hat{T}_b^{-1}=i(-1)^{y+1}\hat{c}_{(x,y+1),\sigma}.
\end{equation}
Therefore, 
\begin{equation}
    \hat{T}_a \hat{\Delta}_{a,b}(\mathbf{r}) \hat{T}_a^{-1}=\hat{\Delta}_{a,b}(\mathbf{r+e_a}),\qquad \hat{T}_a \hat{\Delta}_{c}(\mathbf{r}) \hat{T}_a^{-1}=-\hat{\Delta}_{c}(\mathbf{r+e_a})
\end{equation}
and
\begin{equation}
    \hat{T}_b \hat{\Delta}_{b,c}(\mathbf{r}) \hat{T}_b^{-1}=\hat{\Delta}_{b,c}(\mathbf{r+e_b}),\qquad \hat{T}_b \hat{\Delta}_{a}(\mathbf{r}) \hat{T}_b^{-1}=-\hat{\Delta}_{a}(\mathbf{r+e_b}).
\end{equation}
From Fig.~\ref{Occu_corr}(b), $\langle\hat{\Delta}_{a,b}(\mathbf{r+e_a})\rangle=-\langle\hat{\Delta}_{a,b}(\mathbf{r})\rangle,\quad \langle\hat{\Delta}_{c}(\mathbf{r+e_a})\rangle=\langle\hat{\Delta}_{c}(\mathbf{r})\rangle,\quad \langle\hat{\Delta}_{b,c}(\mathbf{r+e_b})\rangle=-\langle\hat{\Delta}_{b,c}(\mathbf{r})\rangle$ and $\langle\hat{\Delta}_{a}(\mathbf{r+e_b})\rangle=\langle\hat{\Delta}_{a}(\mathbf{r})\rangle$, which demonstrates that pairing order parameter $\langle\hat{\Delta}_\alpha(\mathbf{r})\rangle$ is antisymmetric under both $\hat{T}_a$ and $\hat{T}_b$: $\langle\hat{T}_{a,b}\hat{\Delta}_\alpha(\mathbf{r})\hat{T}_{a,b}^{-1}\rangle=-\langle\hat{\Delta}_\alpha(\mathbf{r})\rangle$.

The phase of the pairing order along bond $\beta$ at position $\mathbf{r}$ with respect to that of a reference bond $\alpha$ at position $\mathbf{r_0}$ is obtained by measuring the phase of the pair-pair correlation: $\text{arg}(\langle \hat{\Delta}^\dagger_\alpha(\mathbf{r}_0)\hat{\Delta}_\beta(\mathbf{r})\rangle)=\text{arg}(\langle\hat{\Delta}_\beta(\mathbf{r})\rangle)-\text{arg}(\langle\hat{\Delta}_\alpha(\mathbf{r_0})\rangle)$.  In Fig. ~\ref{symm_bond}(a-d), we show the patterns of the nearest-neighbor pairing phase using different reference bonds along the $\hat e_a$ or $\hat e_b$ directions. They are found equivalent up to a global phase due to the relative phase of the reference bonds. For example, the phases of the same bond in fig.~\ref{symm_bond}a) and (b)  differ by a constant $\phi_b-\phi_a=-2\pi/3$. Similarly,  $\phi_c-\phi_a=0$, and $ \phi_d-\phi_b=\pi$.

The same pairing symmetry is also observed in 4- and 8-leg systems (see Fig.~\ref{Ne_symm_Y4}(b) and \ref{Y8}(a)). A coexisting CDW is also observed in 4-leg systems with two-holes per stripe (Fig.~\ref{Ne_symm_Y4}(a)), similar to those in the 6-leg ones (Fig.2(b) of the main text). The similar magnitudes of the pairing correlation between different bonds in Fig.~\ref{Y8}(b,c) indicate the isotropy of the SC order.

Fig.~\ref{Sc_long} demonstrates the phases and amplitudes of Cooper pairs beyond nearest neighbors. In Fig.~\ref{Sc_long}(a), we observe a phase change of $-\pi/3$ under bare $C_6$ rotation around $\mathcal{O}$, which is the magnetic $C_6$ rotation in the imaginary $C_{6z}$ gauge (see Fig. 1(a) in the main text). Fig.~\ref{Sc_long}(b)  shows the phase structures  with respect to a different central site (marked by a cross),  which demonstrates $-\pi/3$ phase changes under magnetic $C_6$ rotation ($C_6$ rotation combined with a proper gauge transformation). Pairing is very weak between the two sites related by a site-centered $\pi$ rotation, indicating an odd angular momentum of the SC state~\cite{divic2024anyonsuperconductivitytopologicalcriticality}.

\section{\label{sec: sc_cor}B.  Correlation Functions for $L_y=3-8$}

\begin{figure}
\includegraphics[width=0.9\textwidth,angle=0]{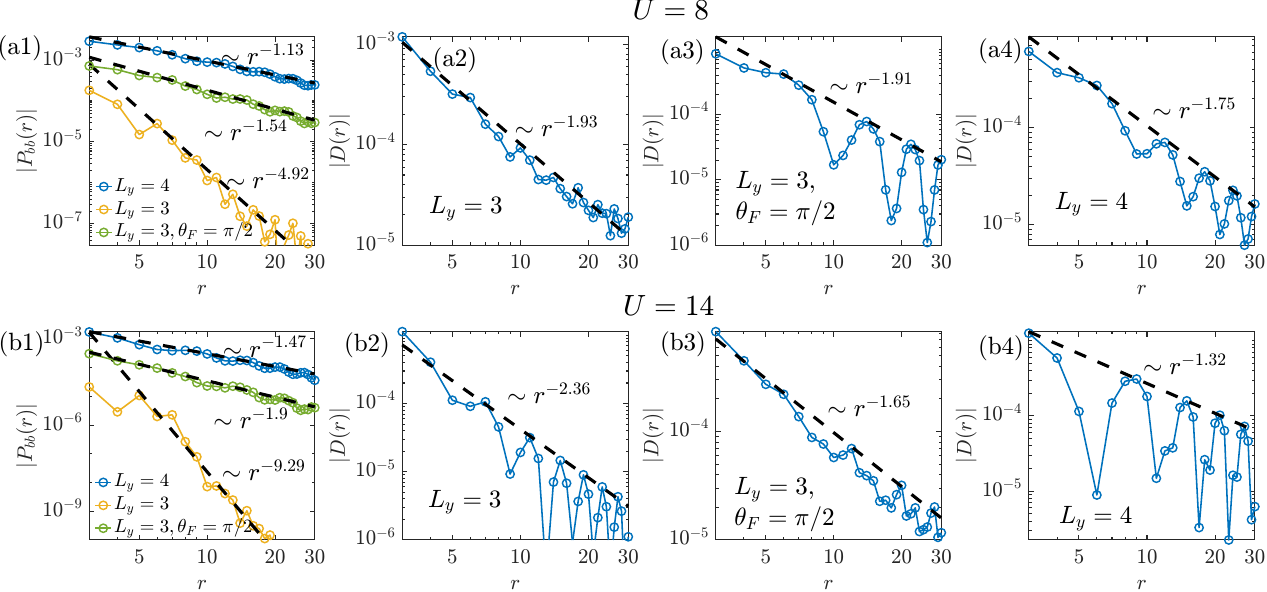}
   \caption{\label{PD_Y34}  (a1-b1) Pair-pair  and (a2-a4, b2-b4) density-density correlations  $P_{bb}(r)$ and $D(r)$ for $U=8, 14$ and $L_y=3, 4$ with or without a charge flux $\theta_F=\pi/2$. Parameters: $L_x=48, \delta=1/12$ and $M=10000$ for $L_y=3$ (12000 for $L_y=4$). Results are converged with the truncation errors smaller than $10^{-8}$. }
\end{figure}

{\bf Pair and density correlations in TSC phase--}The pair-pair and density-density correlations for $L_y=3-4$ are shown  in Fig.~\ref{PD_Y34}. The Landau 
gauge is used for $L_y=3$ as the unit cell in the imaginary $C_{6z}$ gauge does not fit cylinders with odd $L_y$. For $L_y=3$ with periodic boundary condition around the cylinder (y direction),  the density-density correlations dominate the pair-pair ones with the pair Luttinger exponents $K_\text{sc}\gg 2$. Under the insertion of a charge flux $\theta_F=\pi/2$ into the cylinder, however, the pair-pair correlation is strongly enhanced with $K_\text{sc}\lesssim2$.
The $\pi/2$ charge flux makes the odd $L_y$ systems particle-hole symmetric and may be better representatives of the 2D systems~\cite{divic2024chiralspinliquidquantum}. 
For $L_y=4$, the pair correlations are further enhanced and  become dominant, with $K_{sc}=1.13$ and 1.47 for $U=8$ and 14, respectively. 
A summary of the Luttinger exponents for different system sizes with $L_y=3, 4, 6$ is given in Table \ref{tab:Ksc}, which shows that both $K_\text{sc}$ and $K_\text{c}$ get smaller but SC gradually dominates CDW as the system widens.  These results suggest the SC ground state in the 2D  limit.

\begin{table}[htbp]
\centering
\begin{tabular}{ |c|c|c|c|c| } 
\hline
\multirow{2}{*}{$N$} & \multicolumn{2}{|c|}{$K_\text{sc}$} & \multicolumn{2}{|c|}{$K_\text{c}$}\\
\cline{2-5}
                     & $U=8$ & $U=14$& $U=8$ & $U=14$ \\
\hline
$48\times 3$ & 4.92 & 9.29  & 1.93 & 2.36 \\
\hline
$48\times 3, \theta_F=\pi/2$ & 1.54 & 1.9  & 1.91 & 1.65 \\
\hline
$48\times 4$ & 1.13 & 1.47  & 1.75 & 1.32 \\
\hline
$48\times 6$& 0.7& 0.95& 1.58& 1.74\\
\hline
\end{tabular}
\caption{Luttinger exponents $K_\text{sc}$ and $K_\text{c}$ for pair-pair and density-density correlations at different system sizes $N$, interaction strengths $U$ and with periodic boundary condition or with a charge flux  $\theta_F=\pi/2$.}
\label{tab:Ksc}
\end{table}

\begin{figure}
\includegraphics[width=0.95\textwidth,angle=0]{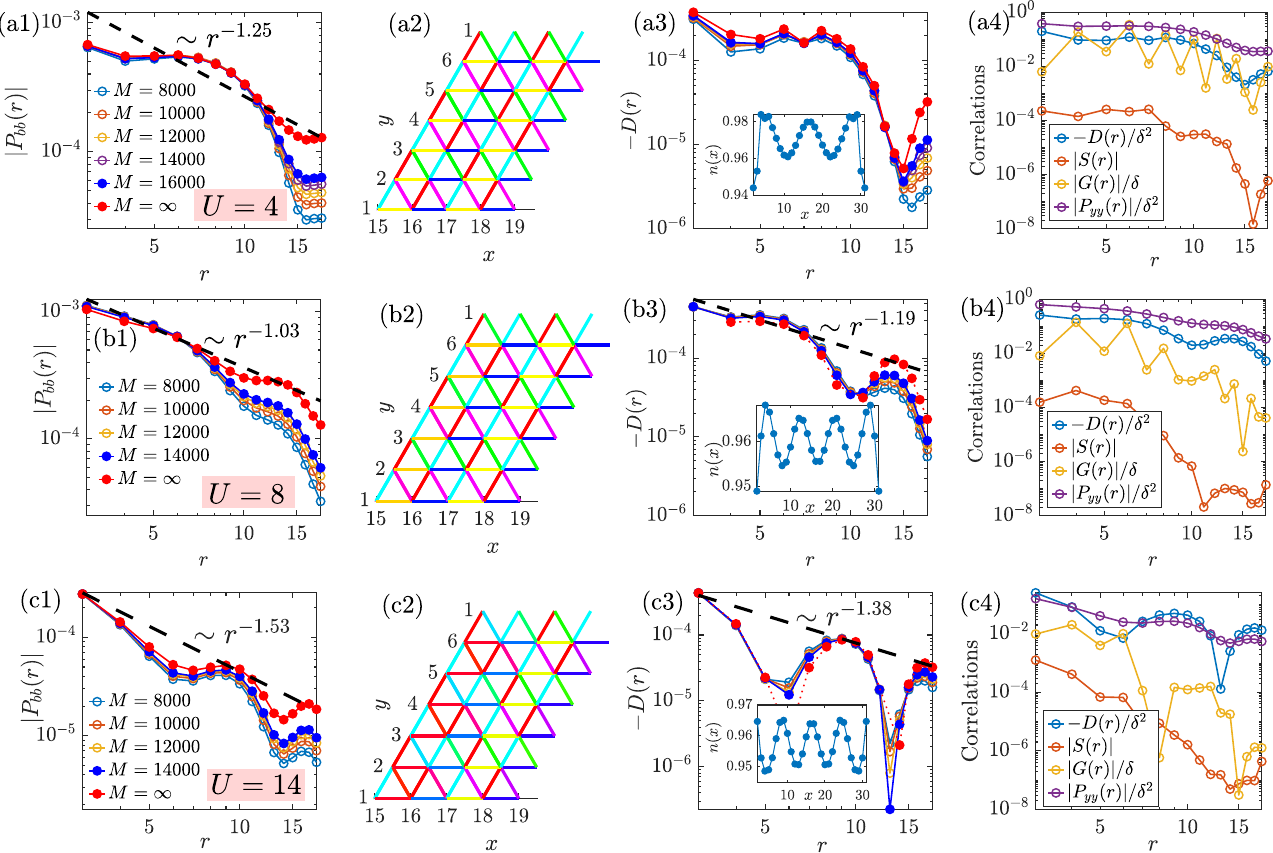}
   \caption{\label{nu24} (a1-c1) Pair-pair correlation; (a2-c2) The pairing phase structure; (a3-c3) Density-density correlation; (a4-c4) Other correlation functions at $M=14000$ for $\delta=1/24$ and $U=4, 8, 14$. The insets in (a3-c3) are the electron density profiles.}
\end{figure}

\begin{figure}
   \includegraphics[width=0.95\textwidth,angle=0]{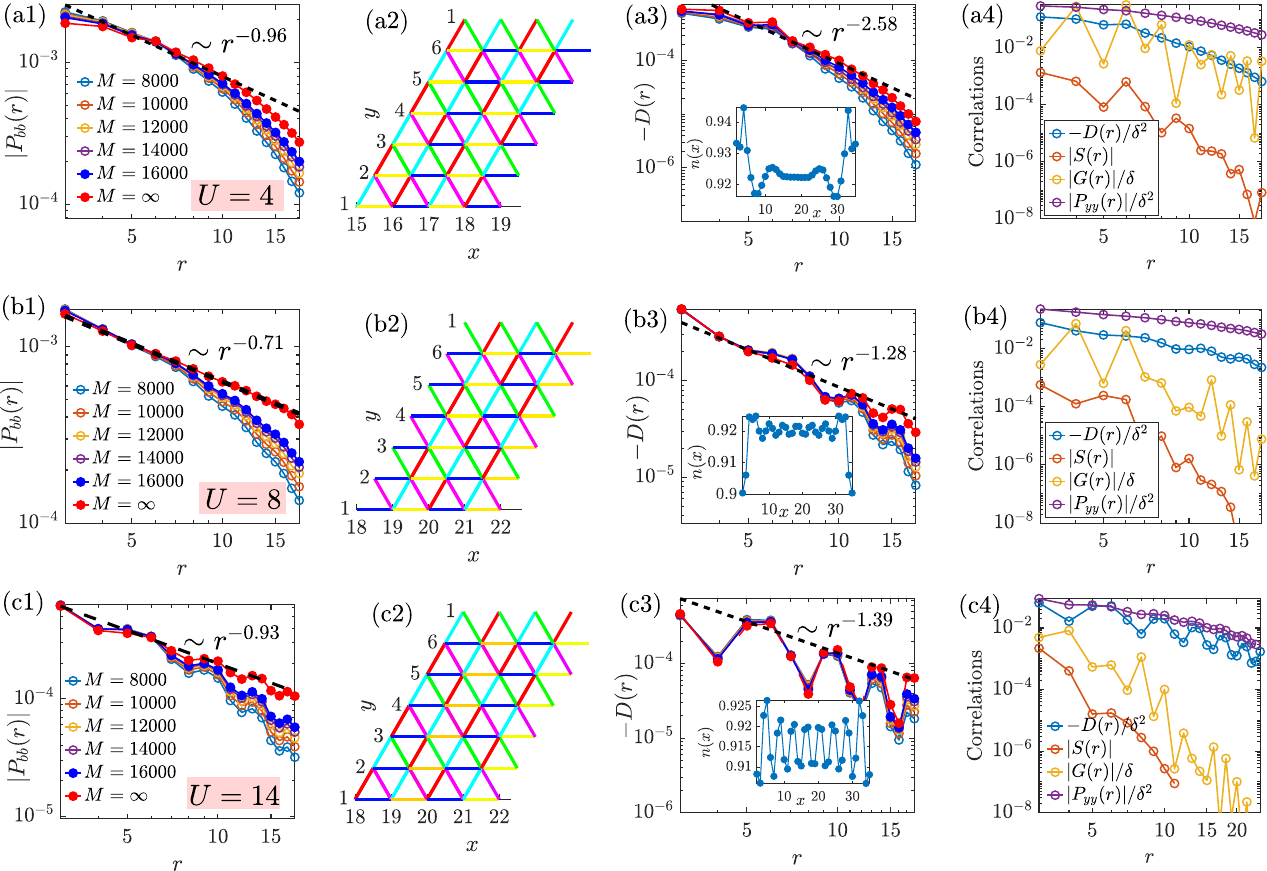}
   \caption{\label{nu12} (a1-c1) Pair-pair correlation; (a2-c2) The pairing phase structure; (a3-c3) Density-density correlation; (a4-c4) Other correlation functions at $M=14000$ for $\delta=1/12$ and $U=4, 8, 14$. The insets in (a3-c3) are the electron density profiles.}
\end{figure}

\begin{figure}
   \includegraphics[width=0.95\textwidth,angle=0]{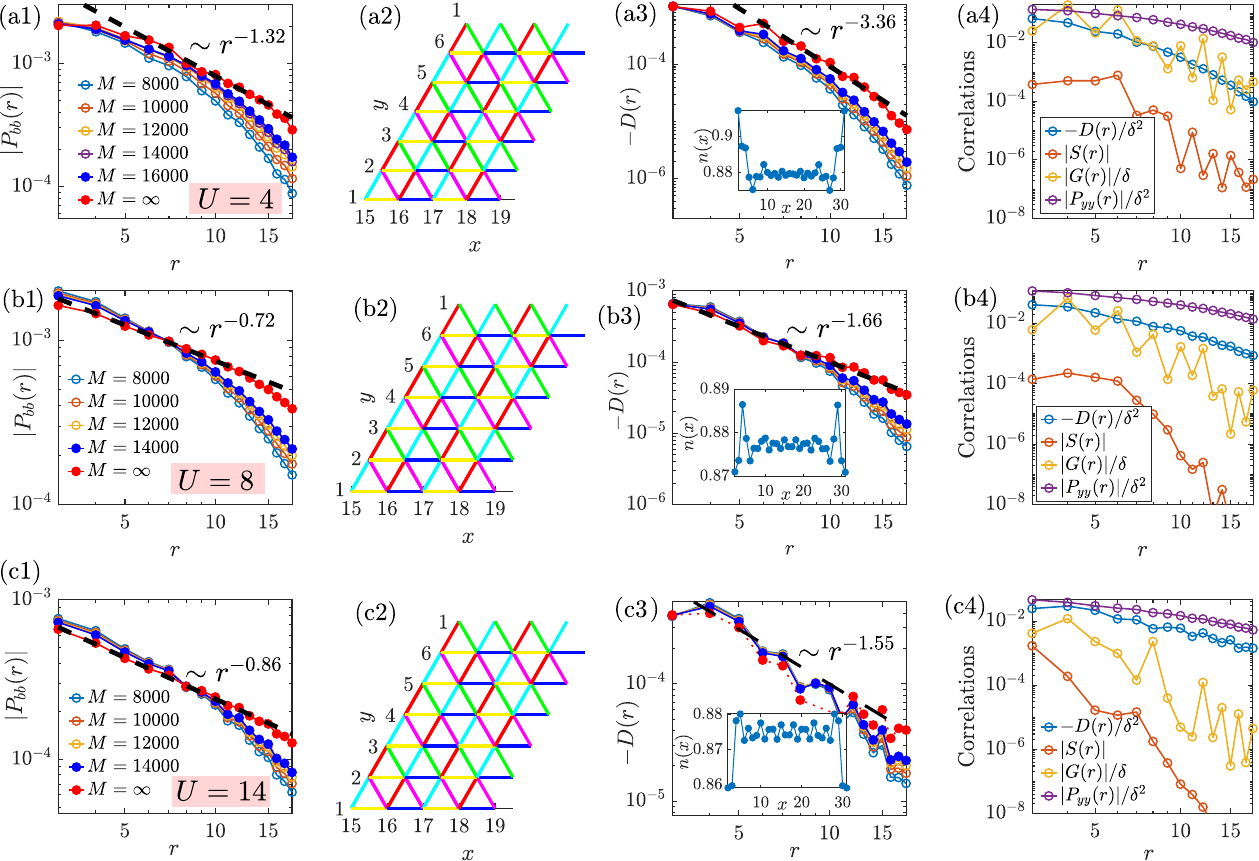}
   \caption{\label{nu8} (a1-c1) Pair-pair correlation; (a2-c2) The pairing phase structure; (a3-c3) Density-density correlation; (a4-c4) Other correlation functions at $M=14000$ for $\delta=1/8$ and $U=4, 8, 14$. The insets in (a3-c3) are the electron density profiles.}
\end{figure}

In Fig.~\ref{nu24}-~\ref{nu8}, we present the pair-pair, density-density and other correlation functions for $\delta=1/24-1/8$, $U=4-14$ and $L_y=6$. The same pairing symmetry  is observed throughout and $|P_{yy}(r)|$ is always much larger than $|G(r)|^2$, indicating a dominant contribution to SC correlation from the charge-2e modes than from the single-particle ones. 
The power-law behavior of pair correlations is established by fitting extrapolated data of $M\rightarrow \infty$
(for smaller $U$, larger bond dimensions upto $M=16000$ data is required for better convergence). 
For slightly larger doping $\nu=1/12$ and $1/8$, we always observe  $K_{SC}<<K_{CDW}$ establishing the dominance of TSC order over CDW order.


\begin{figure}
   \includegraphics[width=0.6\textwidth,angle=0]{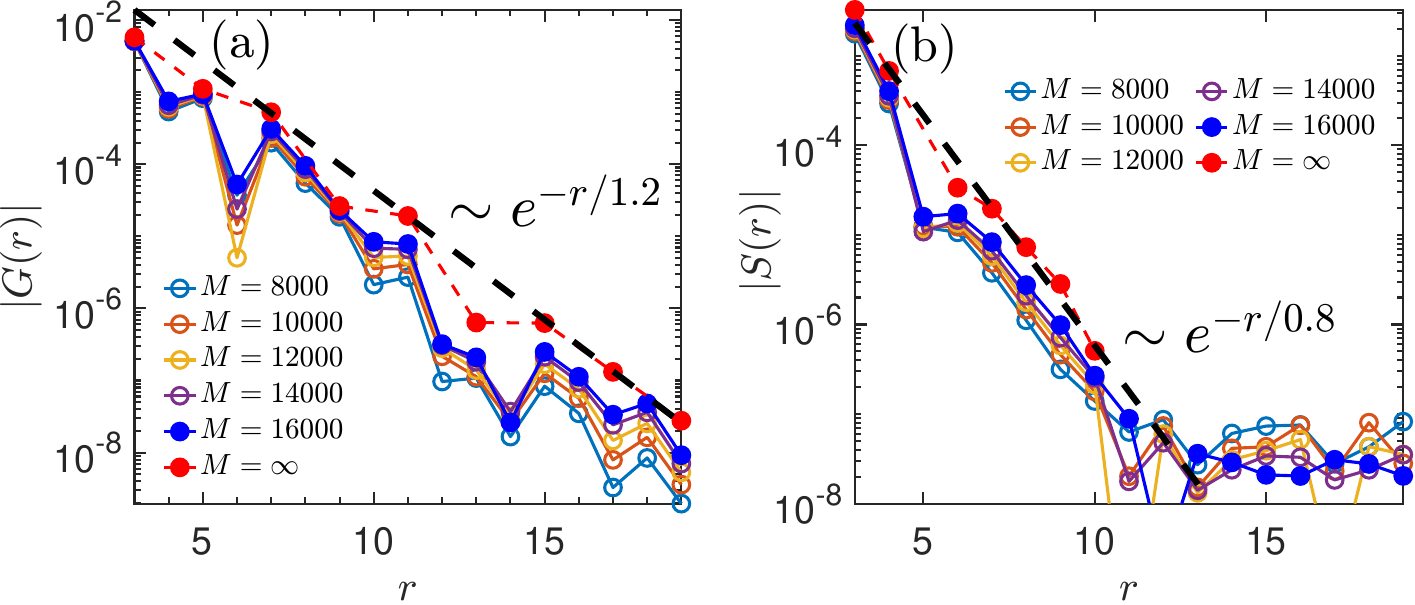}
   \caption{\label{GS} The scalings and fittings of single-particle Green's function $|G(r)|$ and spin correlation $S(r)$ for $U=14$, $N=36\times 6$ and $\delta=1/12$.}
\end{figure}

\begin{figure}
\includegraphics[width=0.4\textwidth,angle=0]{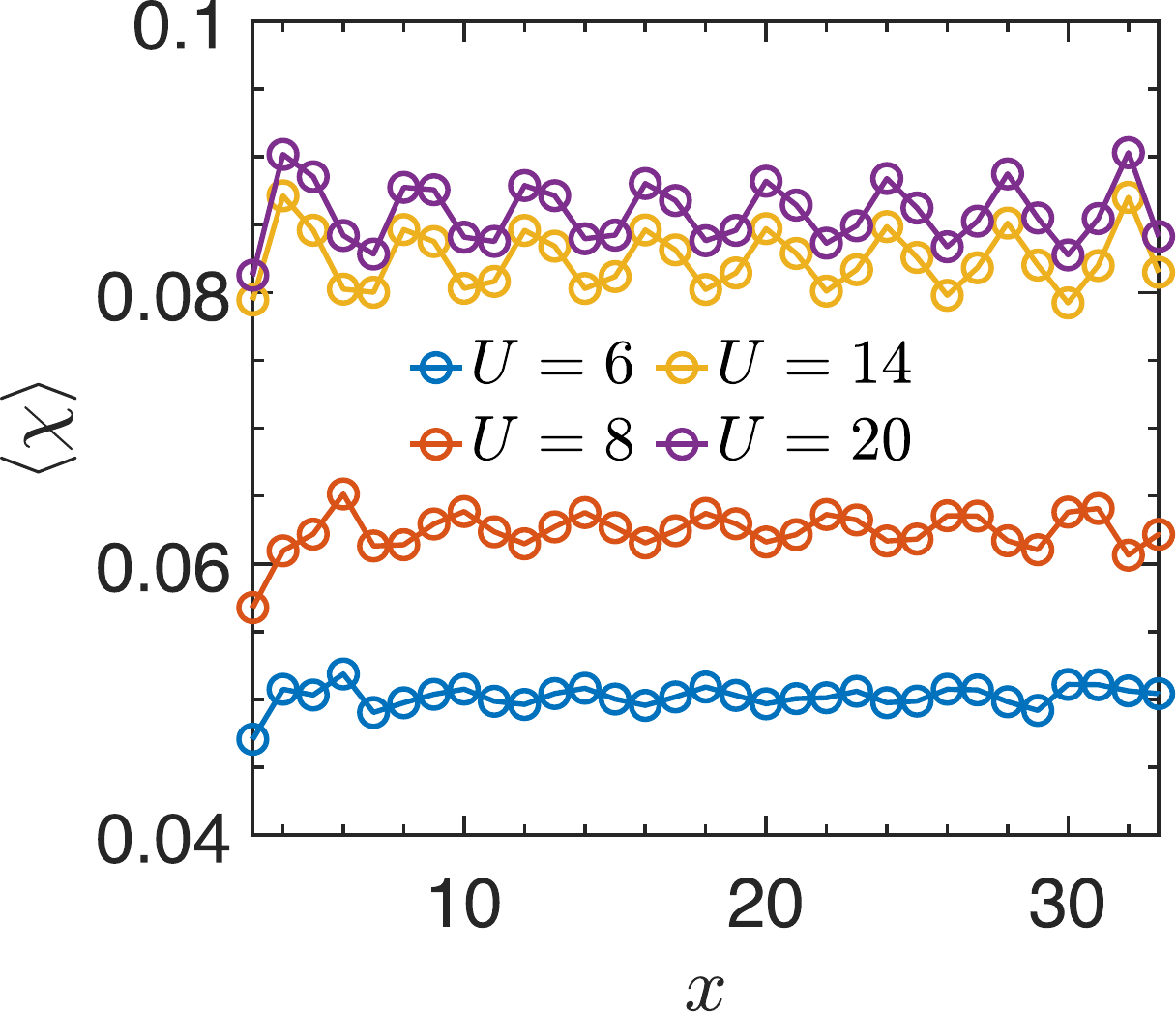}
   \caption{\label{chiral} Spin chiral order $\left \langle \chi \right \rangle = \langle \hat{\boldsymbol{S}}_{i}\cdot (\hat{\boldsymbol{S}}_{j}\times \hat{\boldsymbol{S}}_{k}) \rangle$ for different $U$s, where $i,j,k$ are three sites on the same elementary triangle plaquette in an anticlockwise order. Parameters: $N=36\times 6, \delta=1/12$.}
\end{figure}

\begin{figure}
\includegraphics[width=0.95\textwidth,angle=0]{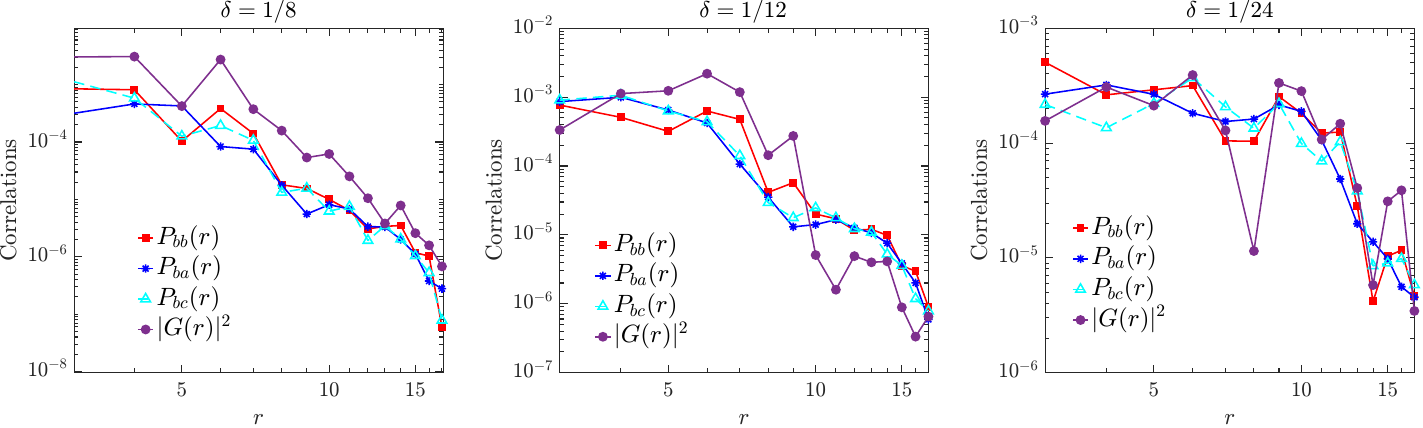}
   \caption{\label{U2}The comparison between pair-pair correlation functions and the single-particle Green's functions squared for $U=2$ at different doping levels.}
\end{figure}

\begin{figure}
   \includegraphics[width=1\textwidth,angle=0]{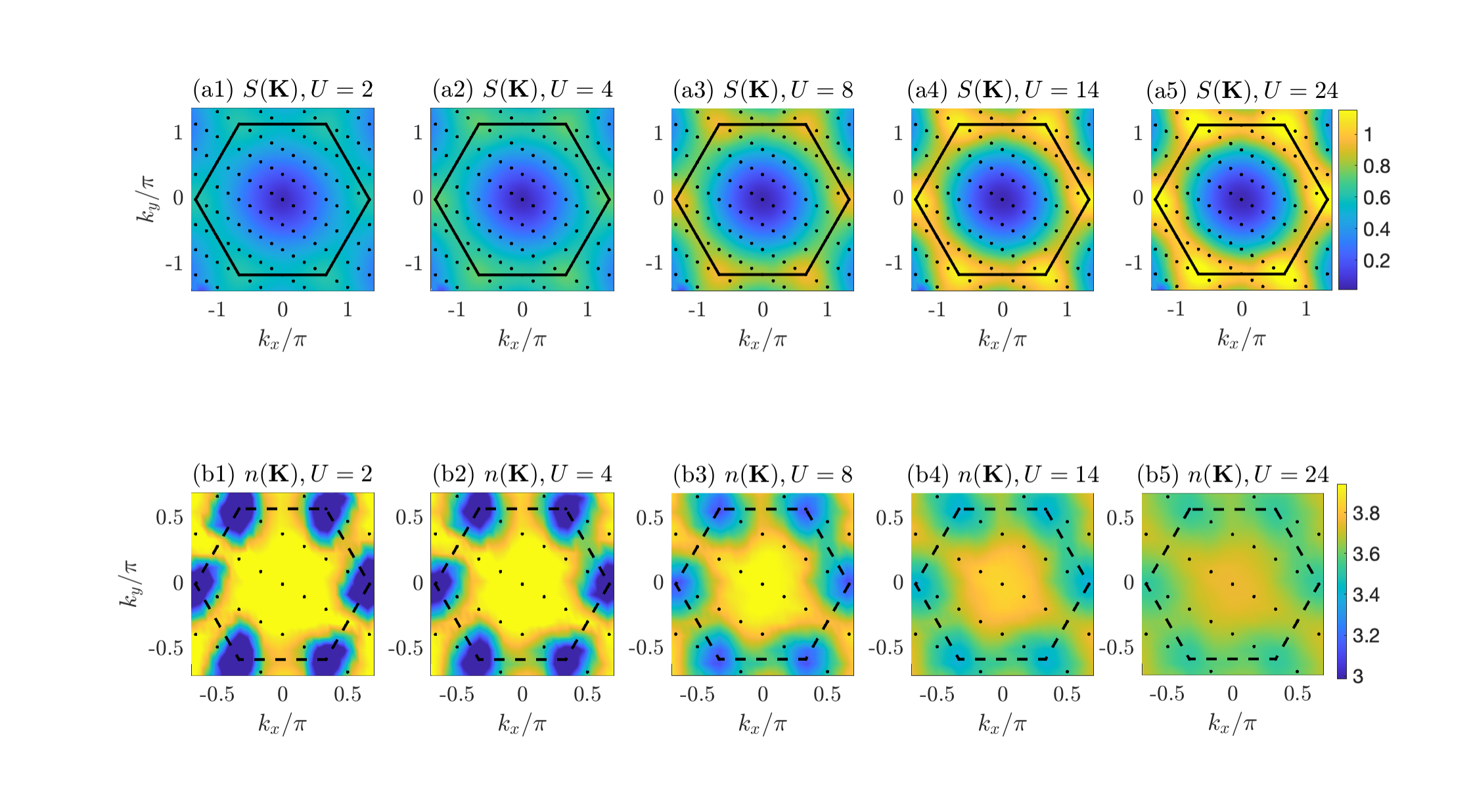}
   \caption{\label{Supp_spinSf} The spin and electron density structure factors $S(\mathbf{k})$ and $n(\mathbf{k})$ in the non-magnetic and magnetic Brillouin zones respectively. Accessible momenta by the cylinder geometry are represented by black dots. Parameters: $N=24\times 6$ and  $\delta=1/12$.}
\end{figure}

 We have also attempted to obtain results on wider $N=48\times 8$ systems. In addition to the same pairing  symmetry  as $L_y=4, 6$ TSC systems (Fig.~\ref{Y8}(a)),  the SC correlations are the dominant ones showing power-law like behavior (Fig. ~\ref{Y8}(b,c)). As $M$ is increased from $18000$ to $24000$, we find much enhanced pair correlations at longer distances. However, much larger $M$ is required to obtain reliable extrapolated data for $M\rightarrow \infty$, which is beyond our current capability. 

{\bf Single-electron, spin and chiral correlations--}We now show some typical results for single-particle Green's function and spin correlations.  Fig.~\ref{GS} shows that both the single-particle Green's function $G(r)=\sum_\sigma\langle \hat{c}^\dagger_\sigma(x_0,y_0) \hat{c}_\sigma(x_0+r,y_0)$ and the spin correlation $S(r)=\langle \hat{\mathbf{S}}(x_0,y_0)\cdot\hat{\mathbf{S}}(x_0+r,y_0)\rangle$ decay exponentially with very short correlation lengths around 1 lattice constant for $L_y=6$ with $U=14$, $\nu=1/12$ and $N=32\times 6$ in a TSC state, indicating finite electron and spin gaps. Similar results are obtained for other  parameters inside the TSC phase.

On the other hand, a finite chiral spin order $\langle \chi \rangle$ is found throughout the TSC phase (Fig.~\ref{chiral} for $\nu=1/12$) and it is stronger at $U$s whose corresponding undoped states are CSL.

{\bf Correlations for smaller $U$ for $L_y=6$ systems--}For a small $U=2$ and doping levels $\nu=1/24-1/8$, we find that the pair correlations decay in a similar manner as the $G^2(r)$ (Fig.~\ref{U2}), indicating a metallic behavior which is consistent with the observed finite charge Chern number $C_c$ shown in the main text as a chiral metal.  However, the finite system we study may have  a temperature cut-off due to the finite dephasing length; thus  we cannot predict the true ground state for zero temperature in thermodynamic limit in such small U regime with very small or vanishing excitation gaps.

\section{\label{sec: structure}C. Electron Occupation Number in Momentum Space and Spin Structure Factor}
Fig.~\ref{Supp_spinSf} shows that as $U$ increases from 2 to 24, corresponding to parent state from deep IQH to deep CSL states, the spin structure factor  $S(\mathbf{k})$ is near featureless for small $U$ and shows peaks 
 at $K$ and $K'$ points in the Brillouin zone for larger $U\sim 8$. Such peaks are further smeared out around the lines connecting $K$ and $K'$ for the intermediate $U$ in the CSL regime
 of half-filling.  

For the electron occupation number in the magnetic Brillouin zone
in the small-$U$ regime, the doped system retains a well-defined Fermi surface  (Fig.~\ref{Supp_spinSf}) indicating an essentially Fermi‑liquid metal with a non‑quantized Hall response
~\cite{PhysRevB.100.115102, MIKou} for the normal state ($T>T_c$). As $U$ increases, the Fermi surface gradually disappears, electronic correlations become dominant, and the TSC appears to emerge directly from a doped CSL.

\section{\label{sec: dqmc1}D. DQMC supplementary data}

\begin{figure}
   \includegraphics[width=\textwidth,angle=0]{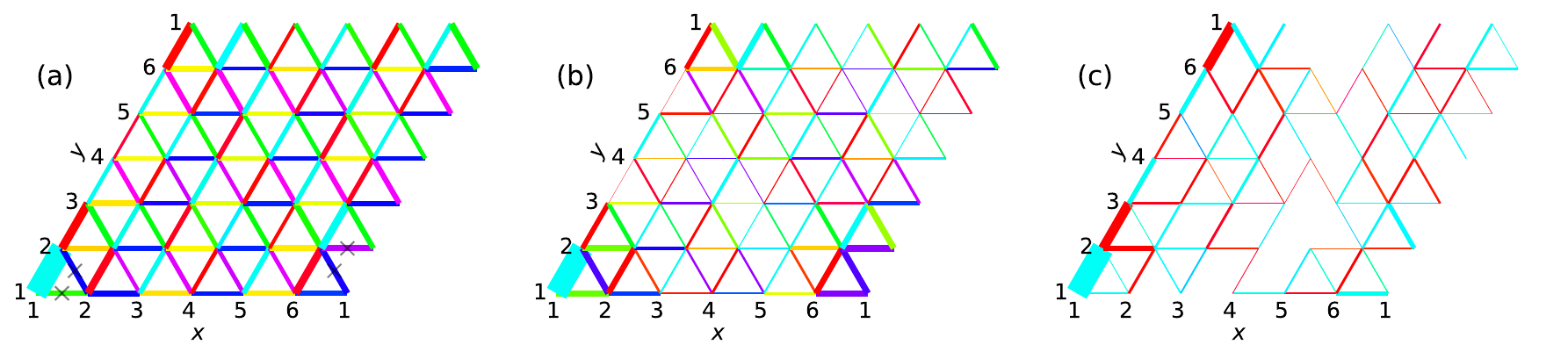}
   \caption{\label{dqmc_phase_comparison} 
   Phase patterns of the DQMC pair–pair correlation at imaginary time $\tau=\beta/2$ for (a) $U/t=8$, $\delta=0.02$ (same data as Fig.\ref{Occu_corr} (b) in the main text); (b) $U/t=2$, $\delta=0.02$; and (c) $U/t=8$, $\delta=0$ (half‑filling). All simulations use a $6\times6$ lattice at $T/t = 1/4$, with the reference bond chosen consistently with the DMRG pattern shown in Fig.~\ref{Occu_corr}(b).
   Line thickness encodes the correlation magnitude; bonds not shown are those for which the jackknife error bar exceeds half the correlation magnitude. Crosses in (a) mark bonds near the reference bond that deviate from the long-range pattern.
   }
\end{figure}

In Fig.~\ref{dqmc_phase_comparison}, we observe that the pairing pattern is disturbed at weak interaction $U/t=2$ or at half-filling, indicating that the pairing pattern is robust at intermediate $U$ and finite doping.
Similar to DMRG, only bonds away from the reference bond preserves the long-range pattern.

\begin{figure}
   \includegraphics[width=\textwidth,angle=0]{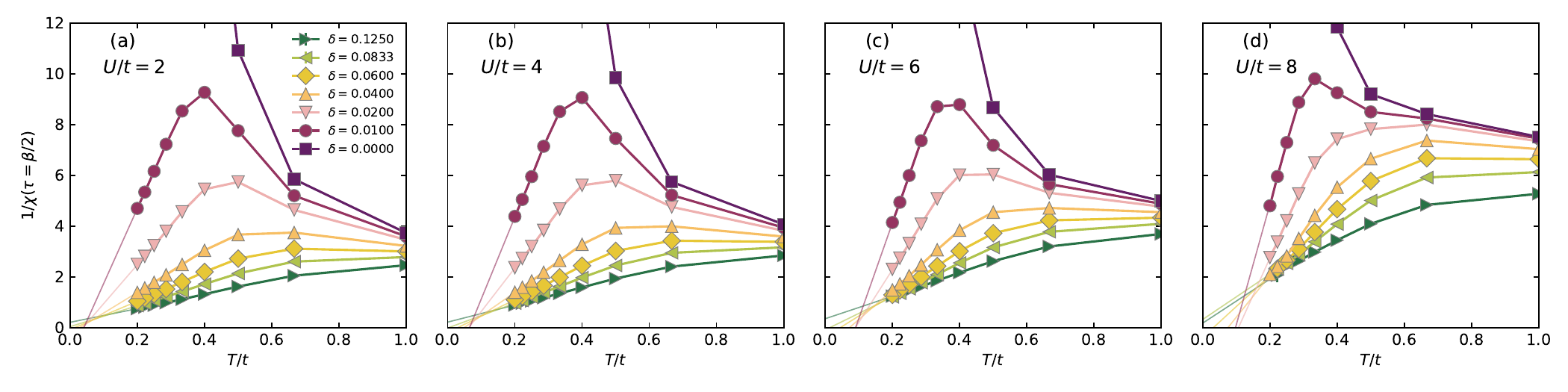}
   \caption{\label{extrapolation} 
Inverse pairing susceptibility $\chi^{-1}(\tau=\beta/2)$ as a function of temperature for different $U$ and $\delta$. A linear extrapolation of $\chi^{-1}(\tau=\beta/2)$ to zero provides an estimate of $T_c$.
   }
\end{figure}
Figure~\ref{extrapolation} displays the temperature dependence of the inverse pairing susceptibility $\chi^{-1}(\tau=\beta/2)$ together with its linear extrapolation to zero, from which the $T_c$ values in Fig.~\ref{Tcvp}(b) are obtained.

\begin{figure}
   \includegraphics[width=\textwidth,angle=0]{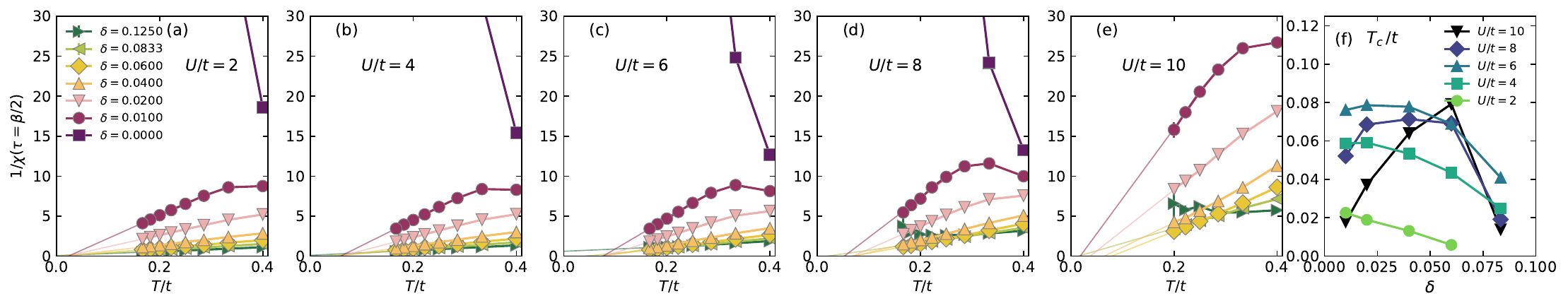}
   \caption{\label{extrapolation_4x4} 
  Temperature dependence of $\chi^{-1}$ and the extrapolated $T_c$ obtained on a $4\times 4$ lattice.
(a)–(e) show analyses similar to Fig.~\ref{extrapolation}, including a larger $U/t=10$. (f) presents the extrapolated $T_c$, in the same manner as Fig.~\ref{Tcvp}.
   }
\end{figure}

\begin{figure}
   \includegraphics[width=\textwidth,angle=0]{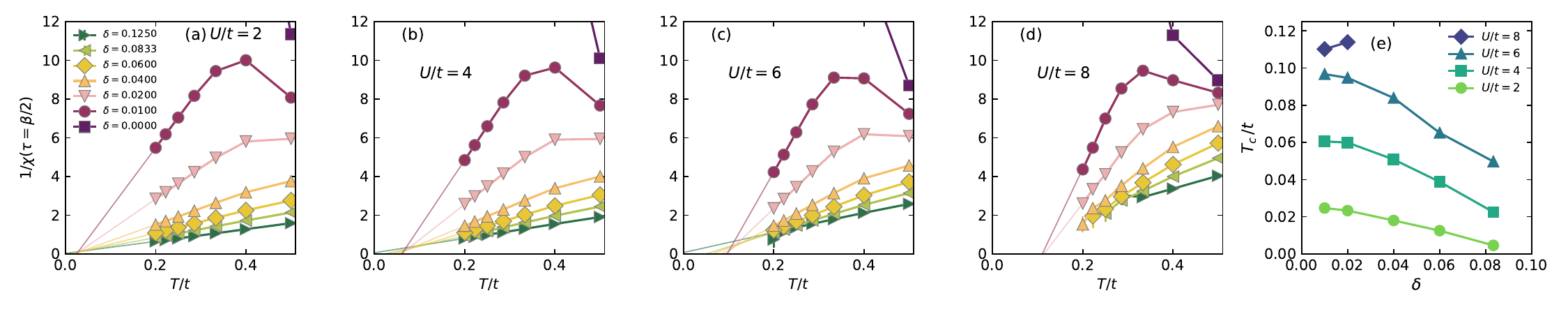}
   \caption{\label{extrapolation_8x8} 
Temperature dependence of $\chi^{-1}$ and extrapolated $T_c$ for an $8\times8$ lattice, displayed in the same layout as Fig.~\ref{extrapolation_4x4}.
   }
\end{figure}

Figure~\ref{extrapolation_4x4} presents the temperature dependence of $\chi^{-1}(\tau=\beta/2)$ on $4\times4$ lattices and the resulting $T_c$ values obtained from its extrapolation. The behavior is qualitatively similar to that on the $6\times6$ lattices, except that the $U$ at which $T_c$ peaks is shifted. Figure \ref{extrapolation_8x8} shows the corresponding results for an $8\times8$ lattice. Although poorer fermion signs restrict the accessible doping, $U$, and temperature ranges, the available data is qualitatively consistent with the $4\times4$ and $6\times6$ systems.

\begin{figure}
   \includegraphics[width=\textwidth,angle=0]{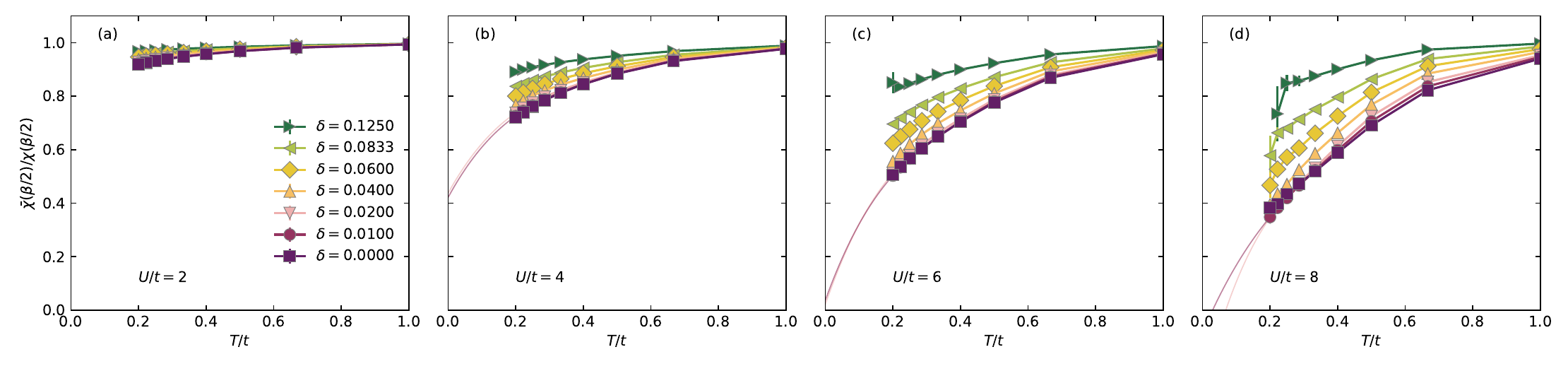}
   \caption{\label{ratio} 
Ratios between the bare (bubble) susceptibility $\bar{\chi}(\tau=\beta/2)$ (constructed from the single‑particle Green’s function) and the full susceptibility $\chi(\tau=\beta/2)$. Simulations are performed on a $6\times6$ lattice. For $\delta=0.01$ and $0.02$, the extrapolation lines are obtained using cubic fits and are shown as visual guides.
   }
\end{figure}
Figure \ref{ratio} plots the ratio $\bar{\chi}(\beta/2)/\chi(\beta/2)$. As expected, it stays near $1$ in the weak‑coupling regime. With increasing $U$, the ratio is strongly suppressed and can even be extrapolated to zero at finite temperature, indicating that the pairing channel is driven by strong correlation effects.

\section{\label{sec: dqmc2}E. Justification of using correlation function at $\texorpdfstring{\tau}{tau}=\texorpdfstring{\beta}{beta}/2$ in DQMC}

In this section, we explain the choice of using $\tau=\beta/2$ in our DQMC correlation functions. The definition of the imaginary-time correlation function is
\begin{align}
  \chi(\tau)/\beta &\equiv \langle \hat O^\dagger(\tau)\,\hat O\rangle
  = \frac{1}{Z}\sum_{m,n}
    \langle n|\hat O^\dagger|m\rangle
    \langle m|\hat O|n\rangle\,
    e^{-\beta E_n}\,e^{\tau(E_n-E_m)},                        \label{eq:chi_tau}
\end{align}
where $Z=\Tr[e^{-\beta(H-\mu N)}]$ is the grand-canonical partition function and
\begin{equation}
  \hat O=\sum_{\alpha,x,y}\Delta_\alpha(x,y)\,e^{-i\theta_\alpha(x,y)} \label{eq:order_parameter}
\end{equation}
is the bosonic order parameter of the pairing channel under study.  The matrix element $O_{mn}\equiv\langle m|\hat O|n\rangle$ connects the eigenstates $|m\rangle$ and $|n\rangle$ of $H-\mu N$ with eigenvalues $E_m$ and $E_n$, respectively.

Because $\hat O$ is bosonic, the Matsubara frequencies are $\omega_n=2\pi n/\beta$ $(n\in\mathbb Z)$. The susceptibility in frequency space reads
\begin{equation}
  \chi(i\omega_n)=\!\int_0^\beta\!\!d\tau\,\chi(\tau)e^{i\omega_n\tau}
  =-\frac{1}{Z}\sum_{m,n}\!
      \frac{|O_{mn}|^{2}\bigl(e^{-\beta E_n}-e^{-\beta E_m}\bigr)}
           {i\omega_n+E_n-E_m}.                                   \label{eq:chi_iwn}
\end{equation}
Analytic continuation $i\omega_n\to\omega+i0^+$ gives the retarded response
\begin{equation}
  \chi(\omega)=
  -\frac{1}{Z}\sum_{m,n}
      \frac{|O_{mn}|^{2}\bigl(e^{-\beta E_n}-e^{-\beta E_m}\bigr)}
           {\omega+E_n-E_m+i0^+}.                                  \label{eq:chi_omega}
\end{equation}

Using $1/(x+i0^+)=\mathcal P(1/x)-i\pi\delta(x)$, the imaginary part is
\begin{equation}
  \Im\chi(\omega)=\frac{\pi}{Z}\sum_{m,n}
     |O_{mn}|^{2}\bigl(e^{-\beta E_n}-e^{-\beta E_m}\bigr)\,
     \delta\!\left(\omega-(E_m-E_n)\right).                       \label{eq:Im_chi}
\end{equation}
Hence $\Im\chi(\omega)$ directly resolves the excitation energies $\omega=E_m-E_n$ that couple to the pairing operator $\hat O$.

Combining Eqs.~\eqref{eq:chi_tau} and \eqref{eq:Im_chi} gives the relationship
\begin{equation}
  \chi(\tau)=\int_{-\infty}^{\infty}\!\beta d\omega\;
     \frac{e^{-\tau\omega}}{1-e^{-\beta\omega}}\,
     \frac{\Im\chi(\omega)}{\pi}.                                 \label{eq:chi_tau_spectral}
\end{equation}
Because $\Im\chi(\omega)$ is odd in $\omega$ (swap $m\leftrightarrow n$ in
\eqref{eq:Im_chi}), we can fold to positive frequencies and write
\begin{equation}
  \chi(\tau)=\int_{0}^{\infty}\!\beta d\omega\;
     W_\tau(\omega)\,
     \frac{\Im\chi(\omega)}{\pi},                          \label{eq:chi_tau_kernel}
\end{equation}
with the effective kernel,
\begin{equation}
  W_\tau(\omega)=\frac{e^{-\tau\omega}+e^{-(\beta-\tau)\omega}}
                      {1-e^{-\beta\omega}}
                =\frac{\cosh\!\bigl[(\beta/2-\tau)\omega\bigr]}
                       {\sinh(\beta\omega/2)}.             \label{eq:kernel}
\end{equation}
Equation~\eqref{eq:kernel} is manifestly symmetric under
$\tau\!\leftrightarrow\!\beta-\tau$ and depends only on the distance
$|\beta/2-\tau|$.  Setting $\tau=\beta/2$ gives
\begin{equation}
  \chi\!\left(\tau=\tfrac{\beta}{2}\right)
  =\int_{0}^{\infty}\!\beta d\omega\;
     \frac{1}{\sinh(\beta\omega/2)}\,
     \frac{\Im\chi(\omega)}{\pi}.                                 \label{eq:chi_beta_half}
\end{equation}
The kernel $[\sinh(\beta\omega/2)]^{-1}$ peaks sharply at $\omega=0$ and decays exponentially once $|\beta\omega|\gtrsim2$.
For a general $\tau$ the low-frequency expansion of $W_\tau(\omega)$ reads
\[
  W_\tau(\omega)=\frac{2}{\beta\omega}
                 \Bigl[1+\frac{(\beta/2-\tau)^2\omega^2}{2}
                 - \frac{\beta^2\omega^2}{24}
                 +\mathcal O(\omega^{4})\Bigr],
\]
so $\tau=\beta/2$ minimises the sub-leading term and therefore assigns the largest relative weight to $|\omega|\lesssim2/\beta$. 

From Eqs.~\eqref{eq:chi_iwn} and \eqref{eq:Im_chi}, the $\omega=0$ susceptibility is
\begin{equation}
  \chi(\omega=0)=\int_{0}^{\infty}\!\beta d\omega\;
     \frac{2}{\beta\omega}\,
     \frac{\Im\chi(\omega)}{\pi},  
     \label{eq:chi_omega_zero}
\end{equation}
whose algebraic kernel $(\beta\omega)^{-1}$ suppresses high-frequency
contributions only as a power law.  Equation~\eqref{eq:chi_beta_half}
shares the same low-frequency $(\beta\omega)^{-1}$ behavior at low frequencies, but damps the high-frequency features exponentially, making $\chi(\tau = \beta/2)$ a more selective probe of low-energy physics.

As the temperature is lowered, the effective low-energy window of
Eq.~\eqref{eq:chi_beta_half} narrows like $1/\beta$, which sharpens its focus on excitations near $\omega=0$.  Although the fermion sign problem prevents access to arbitrarily low temperatures, the $\chi(\tau = \beta/2)$ correlation function offers the best opportunity for
direct comparison with ground-state methods such as DMRG, and provides the most sensitive indicator of the superconducting transition temperature $T_c$.

\section{\label{sec: perturb}F. Pairing Symmetry from Perturbative Approach}
\begin{figure}[t]
    \centering
    \includegraphics[width=0.8\linewidth]{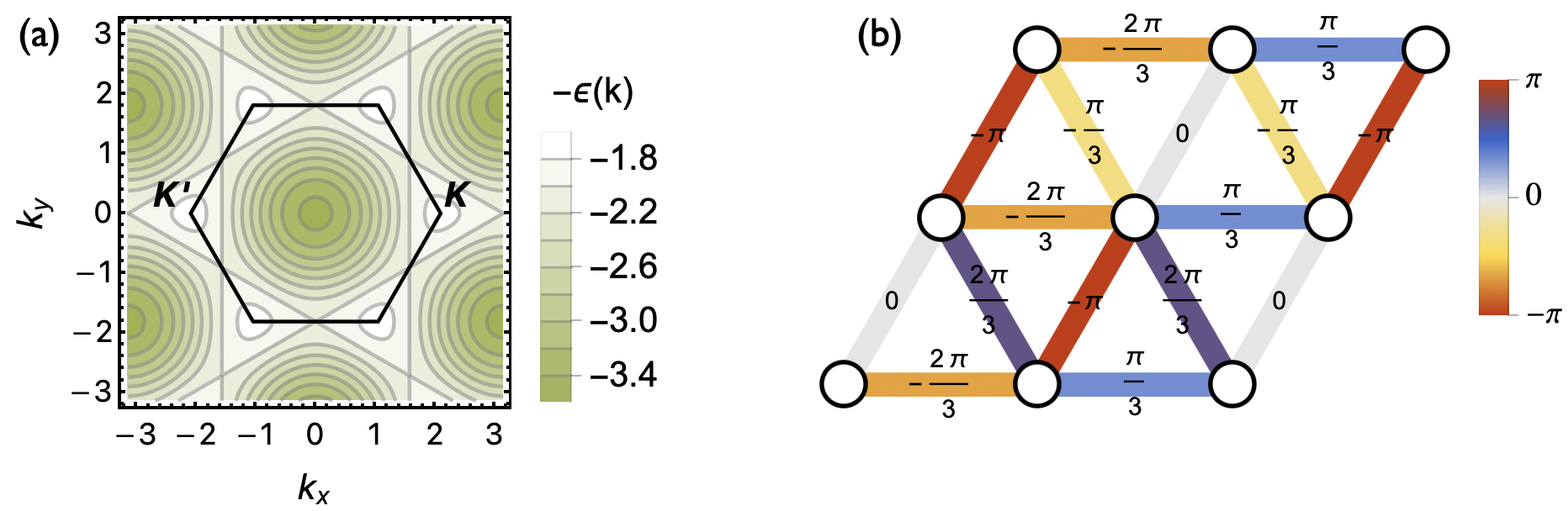}
    \caption{(a) Dispersion of the lower branch of the band structure for $H_0$ at half-filling. The black lines indicate the Brillouin zone boundaries, and the  $K/K'$ points mark the band maxima, which become the relevant valleys upon dilute hole doping. (b)  Pairing symmetry pattern under the imaginary $C_6$ gauge, derived from $\tilde{F}_{\alpha \beta}(d)$  in Eq.~\eqref{Ft}.}
    \label{fig:sym}
\end{figure}

The primitive vectors of the Bravais lattice are given as follows:
\begin{equation}\label{a1a2}
    a_1=(2,0),\;\;\;\; a_2=(1, \sqrt{3})
\end{equation}
and the sublattice coordinates are
\begin{equation}
    d_A=(0,0),\;\;\;\; d_B=(1,0),\;\;\;\; d_C=\left(\frac{1}{2}, \frac{\sqrt{3}}{2}\right),\;\;\;\; d_D=\left(\frac{3}{2}, \frac{\sqrt{3}}{2}\right)
\end{equation}
With the imaginary $C_6$ gauge, the free Hamiltonian is given by:
\begin{eqnarray}
\begin{aligned}
    H_0&=i \sum_{ij} \tau_{i j} c_{i\sigma}^{\dagger} c_{j \sigma}+\text{H. C.}+\mu\sum_{ i}  c_{i \sigma}^{\dagger} c_{i \sigma}\\
    &=\sum_{k \sigma}\left(\sum_{m=-,-'} \xi_{-}(k) c_{k m\sigma}^{\dagger} c_{km \sigma}+\sum_{m=+,+'} \xi_{+}(k) c_{km \sigma}^{\dagger} c_{k m \sigma}\right)
\end{aligned}
\end{eqnarray}
where $\alpha$ is the sublattice index and $m$ is the band index, with energy 
\begin{equation}
    \xi_{\pm}(k) \equiv \pm \epsilon(k)+\mu
\end{equation}
and basis transformation
\begin{equation}
    c_{k\alpha, \sigma}=w_{\alpha \mathrm{m}}(k) c_{km\sigma}
\end{equation}
There are four bands in total: two degenerate bands lying below the Fermi energy (labeled $-$ and $-'$) and two above it ($+$ and $+'$) in the half-filling limit. Upon hole doping, the active bands are $-$ and $-'$, and the relevant valleys are labeled $K$ and $K'$, as shown in Fig.~\ref{fig:sym}~(a). The on-site interaction is given by:
\begin{eqnarray}\label{Hint}
    H_{\text {int }}^{(1)}=\sum_{i \alpha} U n_{i\alpha \uparrow} n_{i \alpha \downarrow}=\sum_{k_1 k_2 k_3 k_4} \sum_\alpha U c_{k_1 \alpha \uparrow}^{\dagger} c_{k_2 \alpha\downarrow}^{\dagger} c_{k_3\alpha \downarrow} c_{k_4\alpha \uparrow} \delta_{k_1+k_2, k_3+k_4},
\end{eqnarray}
where the superscript ``1'' denotes first order in $U$, used here to distinguish it from higher-order corrections.  
Although the bare interaction is repulsive, it can still produce an effective attractive channel via polarization effects from particle-hole excitations, as described by the Kohn-Luttinger mechanism. To show this explicitly, the four-point pair–pair correlation function, expanded to second order in $U$, is given by:
\begin{eqnarray}
\begin{aligned}
    &-\left\langle c_{k \alpha_1 \uparrow} c_{-k\alpha_2 \downarrow} c_{-k^{\prime} \alpha_3 \downarrow}^{\dagger} c_{k^{\prime} \alpha_4 \uparrow}^{\dagger}\right\rangle_{U^2}\\
    =&\sum_{\beta, \beta^{\prime}}\left\langle c_{k \alpha_1 \uparrow} c_{k \beta \uparrow}^{\dagger}\right\rangle_0\left\langle c_{-k \alpha_2 \downarrow} c_{-k_1 \beta^{\prime} \downarrow}^{\dagger}\right\rangle_0\left\langle c_{-k^{\prime} \beta \downarrow} c_{-k^{\prime} \alpha_3\downarrow}^{\dagger}\right\rangle_0
    \left\langle c_{k^{\prime} \beta^{\prime} \uparrow} c_{k^{\prime} \alpha_4 \uparrow}^{\dagger}\right\rangle_0\\
    &\times\left[U \delta_{\beta \beta^{\prime}}-U^2 w_{\beta m}\left(k+k^{\prime}+p\right) w_{\beta^{\prime} m}^*\left(k+k^{\prime}+p\right) w_{\beta^{\prime} n}(p) w_{\beta n}^*(p) \sum_p \frac{n_F\left[\xi_m\left(k+k^{\prime}+p\right)\right]-n_F\left[\xi_n(p)\right]}{\xi_m\left(k+k^{\prime}+p\right)-\xi_n(p)}\right],
\end{aligned}
\end{eqnarray}
where the second term in brackets leads to an additional interaction term:
\begin{eqnarray}\label{Hint2}
    H_{\text {int }}^{(2)}=\sum_{k k^{\prime} \alpha\alpha^{\prime}} c_{k^{\prime} \uparrow \alpha}^{\dagger} c_{-k^{\prime} \downarrow \alpha^{\prime}}^{\dagger} \Gamma_{\left(k^{\prime} \uparrow \alpha\right)\left(-k^{\prime} \downarrow \alpha^{\prime}\right)(-k \downarrow \alpha)\left(k \uparrow \alpha^{\prime}\right)}^{(2)} c_{-k \downarrow \alpha} c_{k \uparrow \alpha^{\prime}}
\end{eqnarray}
with
\begin{eqnarray}
    \begin{aligned}
&\Gamma_{\left(k^{\prime} \uparrow \alpha\right)\left(-k^{\prime} \downarrow \alpha^{\prime}\right)\left(-k \downarrow \alpha\right)\left(k \uparrow, \alpha^{\prime}\right)}^{(2)}=\diag{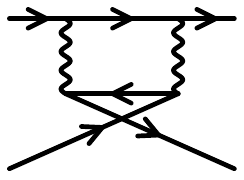}{38pt} \label{f}\\
    =&-U^2 w_{\alpha \mathrm{m}}\left(k+k^{\prime}+p\right) w_{\alpha^{\prime} m}^*\left(k+k^{\prime}+p\right) w_{\alpha^{\prime} n}(p) w_{\alpha \mathrm{n}}^*(p) \times\sum_p \frac{n_F\left[\xi_m\left(k+k^{\prime}+p\right)\right]-n_F\left[\xi_n(p)\right]}{\xi_m\left(k+k^{\prime}+p\right)-\xi_n(p)}
    \end{aligned}
\end{eqnarray}
Note that all other second-order diagrams vanish in the Cooper-pair channel due to the form of the bare vertex in Eq.~\eqref{Hint}. In the band basis, the interaction correction in Eq.~\eqref{Hint2} can be further expressed as:
\begin{eqnarray}\label{Hint3}
    H_{\mathrm{int}}^{(2)}=\sum_{k k^{\prime}}\sum_{m_1 m_2 m_3 m_4} c_{k^{\prime} m_4 \uparrow}^{\dagger} c_{-k^{\prime} m_3 \downarrow}^{\dagger} \delta \Gamma_{\left(k^{\prime} \uparrow m_4\right)\left(-k^{\prime} \downarrow m_3\right)\left(-k_1 m_2\right)\left(k \uparrow m_1\right)} c_{-k m_2 \downarrow} c_{k m_1 \uparrow}
\end{eqnarray}
with
\begin{eqnarray}\label{Gamma2m}
    \delta \Gamma_{\left(k^{\prime} \uparrow m_{4}\right)\left(-k^{\prime} \downarrow m_3\right)\left(-k_{\downarrow} m_2\right)\left(k \uparrow m_1\right)}^{(2)} = w_{\alpha m_{4}}^*\left(k^{\prime}\right) w_{\alpha^{\prime} m_3}^*\left(-k^{\prime}\right) \delta \Gamma_{\left(k^{\prime} \uparrow \alpha\right)\left(-k^{\prime} \downarrow \alpha^{\prime}\right)\left(-k_{ \downarrow} \alpha\right)\left(k \uparrow \alpha^{\prime}\right)}^{(2)} w_{\alpha m_2}(-k) w_{\alpha^{\prime} m_1}(k)
\end{eqnarray}

Assuming dilute hole doping, so that only the \(K\) and \(K'\) valleys are relevant, the basis is given by
\begin{eqnarray}\label{Psi}
    \Psi=\left(\begin{array}{c}
c_{K^{\prime}- \downarrow}  c_{K- \uparrow} \\
c_{K,-\downarrow}  c_{K^{\prime}- \uparrow} \\
c_{K^{\prime}-^{\prime} \downarrow}  c_{K-'\uparrow} \\
c_{K-^{\prime}, \downarrow}  c_{K^{\prime}-^{\prime} \uparrow} \\
c_{K^{\prime}-^{\prime} \downarrow}  c_{K- \uparrow} \\
c_{K^{\prime}- \downarrow}  c_{K-^{\prime} \uparrow} \\
c_{K-^{\prime} \downarrow}  c_{K^{\prime}- \uparrow} \\
c_{K- \downarrow}  c_{K^{\prime}-^{\prime} \uparrow}
\end{array}\right)
\end{eqnarray}
so that the interaction in Eq.~\eqref{Hint3} can be written as an $8\times8$ matrix \(\Psi^{\dagger} \hat{\delta\Gamma} \Psi\). After diagonalizing \(\hat{\delta \Gamma}\), the eigenvalues (in units of \(U^2/t\)) are
\begin{equation}\label{eigenv}
    \boldsymbol{-0.031}, \;\;\;\; 0.014, \;\;\;\;0.023,\;\;\;\; 0.023, \;\;\;\; 0.038, \;\;\;\; 0.041, \;\;\;\;
    0.047, \;\;\;\;
    0.047
\end{equation}
It is important to note that there is only a single negative eigenvalue, indicating a unique pairing channel. The corresponding order parameter is given by:
\begin{equation}\label{Deltach}
    \hat{\Delta}=-c_{K^{\prime}-' \downarrow} c_{K-\uparrow}+c_{K^{\prime}- \downarrow} c_{K-^{\prime} \uparrow}+c_{K-^{\prime} \downarrow} c_{K^{\prime}- \uparrow}-c_{K- \downarrow} c_{K^{\prime}-^{\prime} \uparrow}
\end{equation}
In the following, we determine the pairing symmetry of the order parameter:
\begin{eqnarray}\label{Deltao}
    \begin{aligned}
\hat{\Delta} & =\sum_{\alpha \beta} w_{\alpha -^\prime}^*\left(K^{\prime}\right) w_{\beta -}^*(K)\left(c_{K^{\prime}\alpha \uparrow} c_{K \beta \downarrow}-c_{K^{\prime} \alpha \downarrow} c_{K \beta \uparrow}\right)-\sum_{\alpha \beta} w_{\alpha -^\prime}^*(K) w_{\beta -}^*\left(K^{\prime}\right)\left(c_{K \alpha \uparrow} c_{K^{\prime} \beta \downarrow}-c_{K \alpha \downarrow} c_{K^{\prime} \beta \uparrow}\right)\\
& =\sum_{(\alpha \beta d)} \tilde{F}_{\alpha \beta}(d) \hat{\Delta}_{\alpha \beta}(d)
\end{aligned}
\end{eqnarray}
where \(d\) denotes the relative shift between unit cells for a bond, and \((\alpha \beta d)\) indicates a summation over \(\alpha, \beta,\) and \(d\) without double counting. Additionally, in Eq.~\eqref{Deltao}, we have
\begin{equation}
     \hat{\Delta}_{\alpha \beta}(d)=\frac{1}{N} \sum_{i j}\left(c_{i \alpha \uparrow} c_{i+d \beta \downarrow}-c_{i \alpha \downarrow} c_{i+d \beta \uparrow}\right)
\end{equation}
and
\begin{equation}\label{Ft}
    \tilde{F}_{\alpha \beta}(d)=F_{\alpha \beta}(d)+F_{\beta \alpha}(-d)
\end{equation}
with
\begin{equation}
    F_{\alpha \beta}(d)=w_{\alpha 2}^*\left(K^{\prime}\right) w_{\beta 1}^*(K) e^{-i K d} e^{i K\left(d_\alpha-d_\beta\right)}-w_{\alpha 2}^*(K) w_{\beta 1}^*\left(K^{\prime}\right) e^{i K d} e^{-i K\left(d_\alpha-d_\beta\right)}
\end{equation}
The pairing symmetry can be extracted from \(\tilde{F}_{\alpha \beta}(d)\), which is shown in Fig.~\ref{fig:sym}~(b). The bond phase is invariant under translations by reciprocal lattice vectors defined by the Bravais lattice vectors Eq.~\eqref{a1a2}. Therefore, only a representative portion of the pairing symmetry pattern is shown here, which is in full agreement with the DMRG and DQMC numerical results presented in our work.

We have shown that the second-order perturbation term supports this pairing channel. Next, we examine the effect of the first-order (bare) interaction on this channel. In the basis \(\Psi\) given in Eq.~\eqref{Psi}, the bare interaction matrix is \(U \operatorname{diag}\left(\frac{1}{3} J_2, \frac{1}{3} J_2, \frac{1}{6} J_4\right)\), where \(J_m\) is the \(m\times m\) matrix of ones, and `diag' indicates that these matrices are assembled as block-diagonal components. It is apparent that this matrix vanishes when projected onto the channel given in Eq.~\eqref{Deltach}. This indicates that the effective interaction is entirely given by the second-order vertex \(\Gamma^{(2)}\) in Eq.~\eqref{Gamma2m}, which is negative in this channel, corresponding to an effective attraction.  At the ladder level,
\begin{equation}
  \diag{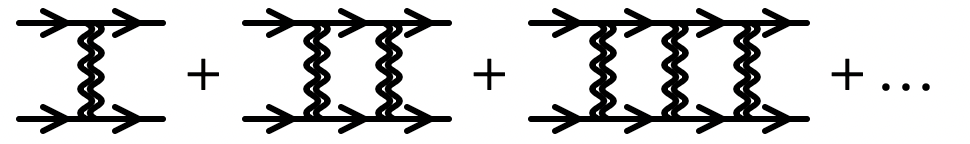}{26pt}  
\end{equation}
this effective attraction (denoted by the double wavy lines) can always lead to an instability toward superconducting order at finite \(T_c\) when there is pairing nesting on the Fermi surface.   

To be more specific, the continuum model can be constructed at the two relevant valleys, with the approximation that the vertex is taken from the half-filled case at zero temperature. Importantly, to preserve all the information about the band-bottom dispersion, we do not use a low-order expansion around the valleys but instead retain the full original band dispersion \(\mathcal{E}_l(q) \equiv \xi_{-}(K_l+q)\). In the basis
\begin{equation}
     \psi_q \equiv\binom{c_{\uparrow}(q)}{c_{\downarrow}^{\dagger}(-q)} \otimes\binom{K}{K^{\prime}} \otimes\binom{-}{-^\prime},
 \end{equation}
the effective Hamiltonian is given by:
\begin{equation}
H_{\mathrm{eff}}=\sum_q \frac{1}{2} \psi_q^{\dagger}\left[\left(\mathcal{E}_K(q)-\mathcal{E}_{K^{\prime}}(q)\right) \sigma_{030}+\left(\mathcal{E}_K(q)+\mathcal{E}_{K^{\prime}}(q)\right) \sigma_{300}\right] \psi_q-g \sum_{qq'}\frac{U^2}{t} \hat{\Delta}^{\dagger}_q \hat{\Delta}_{q^{\prime}}
\end{equation}
where \(g = 0.031\) is the single negative eigenvalue from Eq.~\eqref{eigenv}, and \(\sigma_{abc} = \sigma_a \otimes \sigma_b \otimes \sigma_c\). The SC order parameter is
 \begin{equation}
     \hat{\Delta}_q=\psi_q^{\dagger} \sigma_{122} \psi_q.
 \end{equation}
Since the order parameter anti-commutes with the free Hamiltonian,
 \begin{equation}
     \left\{\sigma_{030}, \sigma_{122}\right\}=0,\;\;\;\;\;\;\;\;\left\{\sigma_{300}, \sigma_{122}\right\}=0,
 \end{equation}
the superconducting gap is fully opened, maximizing kinetic-energy gain. Because the order parameter \(\hat{\Delta}\) corresponds to the most unstable channel in the vertex (with the only negative eigenvalue) and leads to a fully opened gap, it is justified to neglect any detailed \(q\)-dependence within this Cooper-pair channel. Therefore, we perform the following self-consistent calculation using a \(q\)-independent mean-field ansatz \(\langle\hat{\Delta}\rangle = \frac{t \Delta}{g U^2}\). The resulting SC order parameter from the self-consistent calculation is approximately given by:
\begin{equation}
     \Delta \propto \exp \left[-\frac{t}{g U^2 m^*}\right],
 \end{equation}
where \(m^*\) denotes the effective mass.

\end{document}